\documentclass[preprint2]{exoplanet}
\usepackage{amsmath}
\usepackage{times}

\def\apjl{Astrophys.\ J. }
\def\apjs{Astrophys.\ J.\ (Supp.) }
\def\aap{Astron.\ Astrophys. }

\def\figpath{}

\def\Frac#1#2{{{\displaystyle\strut#1}\over{\displaystyle\strut#2}}}

\def\crm{\cr\noalign{\medskip}}
\def\m@th{\mathsurround=0pt}
\def\EQM#1{\vcenter{\normalbaselines\m@th
    \ialign{${\displaystyle ##}$\hfil&&\ ${\displaystyle ##}$\hfil\crcr
    \mathstrut\crcr\noalign{\kern-\baselineskip}
    \noalign{\smallskip}
    #1\crcr\mathstrut\crcr\noalign{\kern-\baselineskip}}}}
\newcommand{\vv}[1]{\boldsymbol{#1}}

\def\aa{\frac{21}{2}}
\def\ab{-}
\def\ac{+9}

\def\sign{\mathrm{sign}}
\def\dw{\delta}
\def \be  {\begin{equation}}
\def \ee  {\end{equation}}
\def \ei {\mathrm{e}}

\def \iii {\mathrm{i}}
\def \ii {\mathrm{I}}
\def \jj {\mathrm{J}}
\def \kk {\mathrm{K}}
\def \Kn {K n}
\def \II {\mathrm{i}}
\def \JJ {\mathrm{j}}
\def \KK {\mathrm{k}}
\def \md {\arrowvert}
\def \ve {\varepsilon}
\def \vf {v}

\def \XX {{\cal E}}

\def \cU {U}
\def \cV {V}
\def \cF {F}
\def \pt {P}
\def \mmu {x}

\def \llabel#1{\label{#1}}

\begin{document}

\title{\textbf{\LARGE Tidal evolution of exoplanets}}

\author {\textbf{\large Alexandre C.M. Correia}}
\affil{\small\em University of Aveiro}

\author {\textbf{\large Jacques Laskar}}
\affil{\small\em Paris Observatory}

\begin{abstract}
\begin{list}{ } {\rightmargin 1in}
\baselineskip = 11pt
\parindent=1pc
{\small 
Tidal effects arise from differential and inelastic deformation of a planet
by a perturbing body. 
The continuous action of tides modify the rotation of the planet together with
its orbit until an equilibrium situation is reached.
It is often believed that synchronous motion is the most probable outcome of
the tidal evolution process, since synchronous rotation is observed for the
majority of the satellites in the Solar System.
However, in the XIX$^\mathrm{th}$ century, Schiaparelli also assumed synchronous
motion for the rotations of Mercury and Venus, and was later shown to be
wrong.
Rather, for planets 
in eccentric orbits synchronous rotation is very unlikely.
The rotation period and axial tilt of exoplanets is still unknown, but
a large number of planets have been detected close to the parent 
star and should have evolved to a final equilibrium situation.
Therefore, based on the Solar System well studied cases, we can make some
predictions for exoplanets.
Here we describe in detail the main tidal effects that modify the
secular evolution of the spin and the orbit of a planet. 
We then apply our knowledge acquired from Solar System situations to exoplanet
cases.
In particular, we will focus on two classes of planets, ``Hot-Jupiters''
(fluid) and ``Super-Earths'' (rocky with atmosphere).
 \\~\\~\\~}
 
\end{list}
\end{abstract}

\section{INTRODUCTION}


The occurrence, on most open ocean coasts, of high sea tide at about the time of
Moon's passage across the meridian, early prompted the idea that Earth's satellite
exerts an attraction on the water. 
The occurrence of a second high tide
when the Moon is on the opposite meridian was a great puzzle, but
the correct explanation of the tidal phenomena was given by Newton
in ``{\it Philosophi{\ae} Naturalis Principia Mathematica}''.
Tides are a consequence of the lunar and solar gravitational forces acting in
accordance with laws of mechanics.
Newton realized that the tidal forces also must affect the
atmosphere, but he assumed
that the atmospheric tides would be too small to be detected, because
changes in weather would introduce large irregular variations upon barometric
measurements.

However, the semi-diurnal oscillations of the atmospheric surface pressure has
proven to be one of the most regular of all meteorological phenomena. 
It is readily detectable by harmonic analysis at any station over the world
\citep[e.g.][]{Chapman_Lindzen_1970}. 
The main difference in respect to ocean tides is that atmospheric tides follow
the Sun and not the Moon,
as the atmosphere is essentially excited by the Solar heat.
Even though tides of gravitational origin are present in the atmosphere, the
thermal tides are more important as the pressure variations on the ground are more
sensitive to the temperature gradients than to the gravitational ones.

The inner planets of the Solar System as well as the majority of the main satellites
present today a spin different from what is believed to have been the initial
one \citep[e.g.][]{Goldreich_Soter_1966, Goldreich_Peale_1968}.
Planets and satellites are supposed to rotate much faster in the beginning and 
any orientation of the spin axis may be allowed
\citep[e.g.][]{Dones_Tremaine_1993,Kokubo_Ida_2007}. 
However, tidal dissipation within the internal layers give rise to secular
evolution of planetary spins and orbits.  
In the case of the satellites, spin and orbital evolution is mainly driven by tidal
interactions with the central planet, whereas for the inner planets the main
source of tidal dissipation is the Sun (in the case of the Earth,
tides raised by the Moon are also important).

Orbital and spin evolution cannot be dissociated because the total
angular momentum must be conserved.
As a consequence a reduction in the rotation rate of a body implies an increment
of the orbit semi-major axis and vice-versa.
For instance, the Earth's rotation period is increasing about 2~ms/century
\citep[e.g.][]{Williams_1990}, and the Moon is consequently moving away about
3.8~cm/year \citep[e.g.][]{Dickey_etal_1994}.
On the other hand,
Neptune's moon, Triton, and the Martian moon, Phobos, are spiraling down into the
planet, clearly indicating that the present orbits are not primordial, and may
have undergone a long evolving process from a previous capture from an
heliocentric orbit \citep[e.g.][]{Mignard_1981m,Goldreich_etal_1989,Correia_2009}.
Both the Earth's Moon and Pluto's moon, Charon, have a significant fraction of the mass
of their systems, and therefore they could be classified as double-planets rather than
as satellites. 
The proto-planetary disk is unlikely to produce double-planet systems, whose
origin seems to be due to a catastrophic impact of the initial planet with a body of
comparable dimensions \cite[e.g.][]{Canup_Asphaug_2001,Canup_2005}.
The resulting orbits after collision are most likely eccentric, but the present
orbits are almost circular suggesting that 
tidal evolution subsequently occurred.

The ultimate stage for tidal evolution corresponds to the synchronous rotation, a configuration
where the rotation rate coincides with the orbital mean motion, since synchronous
equilibrium corresponds to the minimum of dissipation of energy. 
However, when the eccentricity is different from zero
some other configurations are possible, such as the 3/2 spin-orbit resonance
observed for planet Mercury
\citep[][]{Colombo_1965,Goldreich_Peale_1966,Correia_Laskar_2004} or the
chaotic rotation of Hyperion \citep[][]{Wisdom_etal_1984}.
When a dense atmosphere is present, thermal atmospheric tides may also
counterbalance the gravitational tidal effect and non-resonant equilibrium configurations are
also possible, as it is illustrated by the retrograde rotation of Venus
\citep{Correia_Laskar_2001}.
Additional effects may also contribute to the final evolution of the spin,
such as planetary perturbations or core-mantle friction.

Despite the proximity of Mercury and Venus to the Earth, the determination of
their rotational periods has only been achieved in the second half of the 
XX$^\mathrm{th}$ century, when it became possible to use radar ranging on the
planets \citep{Pettengill_Dyce_1965, Goldstein_1964, Carpenter_1964}. 
We thus do not expect that it will be easy
to observe the rotation of the recently discovered exoplanets.
Nevertheless, many of the exoplanets are close to their host star, and we can
assume that exoplanets' spin and orbit have already undergone enough dissipation and
evolved into a final equilibrium possibility.
An identical assumption has been done before by Schiaparelli for Mercury and Venus 
(\citeyear{Schiaparelli_1889}), who made predictions for their
rotations based on Darwin' work (\citeyear{Darwin_1880}).
Schiaparelli's predictions were later proved to be wrong, but were nevertheless much
closer to the true rotation periods than most values derived form observations
in the two previous centuries. 
As Schiaparelli, we may dare to establish predictions for
the rotation periods of some already known exoplanets. 
We hope that the additional knowledge that we gained from a better understanding
of the rotation of Mercury and Venus will help us to be at least as close to the
reality as Schiaparelli was. 
Indeed, observations also show that many of the exoplanets have
very eccentric orbits. 
In some cases eccentricities larger than 0.9 are found
\citep[e.g.][]{Naef_etal_2001,Jones_etal_2006,Tamuz_etal_2008}, which opens a
wide variety for final tidal equilibrium positions, different from what we
observe around the Sun.


In this Chapter we will describe the tidal effects that modify the
secular evolution of the spin and orbit of a planet. 
We then apply our knowledge acquired from Solar System situations to exoplanet
cases.
In particular, we will focus on two classes of planets, ``Hot-Jupiters''
(fluid) and ``Super-Earths'' (rocky), which are close to the star and
therefore more susceptible of being arrived in a final equilibrium situation.

\section{MODEL DESCRIPTION}

We will first omit the tidal effects, and describe the spin motion of the
planet in a conservative framework.
The motion equations will be obtained from an Hamiltonian formalism
\citep[e.g.][]{Goldstein_1950}
of the total gravitational energy of the planet (Sect.~2.1).
Gravitational tides (Sect.~2.2) and thermal atmospheric tides (Sect.~2.3) will
be described later.
We also discuss the impact of spin-orbit resonances (Sect.~2.4) and planetary
perturbations (Sect.~2.5).

\subsection{Conservative motion}

The planet is considered here as a rigid body with mass $ m $ and 
moments of inertia $ A \le B < C $, supported by the reference frame $ ( \vv{\ii},
\vv{\jj}, \vv{\kk} ) $, fixed with respect to the planet's figure.
Let $ \vv{L} $ be the total rotational angular momentum and $ ( \vv{\II},
\vv{\JJ},\vv{\KK} ) $ a  reference frame linked to the orbital plane
(where $ \vv{\KK} $ is the normal to this plane). 
As we are interested in the long-term behavior of the spin axis, 
we merge the axis of figure $\vv{\kk}$ with direction of the angular momentum
$ \vv{L} $. 
Indeed, the average of $ \vv{\kk} $ coincides with 
$ \vv{L} / L $ up to $J^2$, where $ \cos J = \vv{L} \cdot \vv{\kk} $
\citep{Boue_Laskar_2006}.
$J$ is extremely small for large rocky planets ($J\approx 7 \times 10^{-7} $
for the Earth), being even smaller for Jupiter-like planets that behave as fluids.
The angle between $ \vv{\kk} $ and $ \vv{\KK} $ is the obliquity, $ \ve $, and
thus, $ \cos \ve = \vv{\KK} \cdot \vv{\kk} $ (Fig.\,\ref{Fig01}).

\begin{figure}[t]
 \epsscale{1.}
\plotone{\figpath 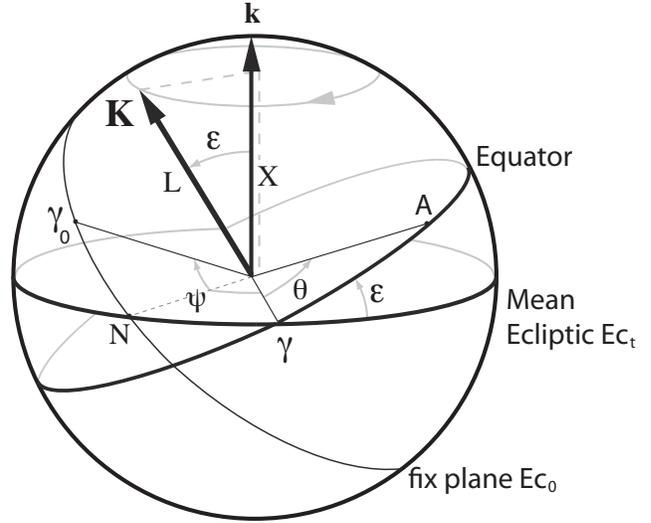}
  \caption{\small Andoyer's canonical variables. $ L $ is the projection of the total
    rotational angular momentum vector $ \vv{L} $ on the principal axis of
    inertia $ \vv{\kk} $, and $ X $ the projection of the angular momentum
    vector on the normal to the orbit (or ecliptic) $ \vv{\KK} $. The 
    angle between the equinox of date $ \gamma $ and a fixed point of the
    equator $ A $ is the hour angle $ \theta $, and $ \psi = \gamma $ N + N $
    \gamma_0 $ is the general precession angle. The direction of $ \gamma_0 $ is on
    a fixed plane $ {E_c}_0 $, while $ \gamma $ is on the mean orbital (or
    ecliptic) $ {E_c}_t $ of date $ t $. \llabel{Fig01}}  
 \end{figure}

The Hamiltonian of the motion can be written
using canonical Andoyer's action variables ($ L , X $) and their conjugate
angles ($ \theta , - \psi $) \citep{Andoyer_1923,Kinoshita_1977}.
$ L = \vv{L} \cdot \vv{\kk} = C \omega $ is the projection of the angular
momentum on the $ C $ axis, with rotation rate $ \omega = \dot \theta - \dot \psi
\cos \ve $, and $ X = \vv{L}
\cdot \vv{\KK} $ is the projection of the angular momentum on the normal to the ecliptic; 
$ \theta $ is the hour angle between the equinox of date and a fixed point of
the equator, and $ \psi $ is the general precession angle, an angle that
simultaneously accounts for the precession of the spin axis and the orbit
(Fig.\,\ref{Fig01}).

\subsubsection{Gravitational potential}

\llabel{030214c}

The gravitational potential $ \cV $ (energy per unit mass) generated by the
planet at a generic point of the space $ \vv{r} $, expanded in degree two of
$R/r$, where $R$ is the planet's radius, is given by \citep[e.g.][]{Tisserand_1891,Smart_1953}:  
\begin{eqnarray}
\cV (\vv{r}) = - \Frac{G m}{r} & + & \Frac{G
(B - A)}{r^3} P_2 ( \vv{\hat r} \cdot \vv{\jj} ) \crm & + &  \Frac{G (C - A)}{r^3}
P_2 ( \vv{\hat r} \cdot \vv{\kk} ) \ , \llabel{090418a}
\end{eqnarray}
where  $ \vv{\hat r} = \vv{r} / r $, $ G $ is the gravitational constant, and $
P_2 (x) = (3 x^2 -1)/2 $ are the Legendre polynomials of degree two.
The potential energy $ \cU $ when orbiting a central star of mass $ m_\star $ 
is then:
\be
\cU = m_\star \cV (\vv{r}_{}) \llabel{021209a} \ .
\ee

For a planet evolving in a
non-perturbed keplerian orbit, we write:
\be
\vv{\hat r} = \cos (\varpi + \vf) \vv{\mathrm{\II}} + \sin (\varpi + \vf)
\vv{\mathrm{\JJ}} \ , \llabel{090418b}
\ee
where $ \varpi $ is the longitude of the periapse and $ \vf $ the true anomaly
(see Chapter~2: {\it Keplerian Orbits and Dynamics}).
Thus, transforming the body equatorial frame $ ( \vv{\ii}, \vv{\jj}, \vv{\kk} )
$ into the orbital frame $ ( \vv{\II}, \vv{\JJ}, \vv{\KK} ) $, we obtain
(Fig.\,\ref{Fig01}): 
\be
 \left\{ 
  \begin{array}{l l}
   \vv{\hat r} \cdot \vv{\jj}  = - \cos w \sin \theta + \sin w \cos \theta \cos \ve 
     \ , \crm
   \vv{\hat r} \cdot \vv{\kk}  = - \sin w \sin \ve \ ,
  \end{array} 
 \right. \llabel{090418c}
\ee 
where $ w = \varpi + \psi + \vf $ is the true longitude of date.
The expression for the potential energy (Eq.\,\ref{021209a}) becomes
\citep[e.g.][]{Correia_2006}:
\begin{eqnarray} 
\cU & = & - \Frac{G m m_\star}{r_{}} + \Frac{G C m_\star}{r_{}^3} E_d P_2 ( \sin
w \sin \ve ) \crm & & - \Frac{3 G m_\star}{8 r_{}^3} (B - A) \, F (\theta, w, \ve)
\ , \llabel{030123b} 
\end{eqnarray}  
where
\begin{eqnarray} 
& F (\theta, w, \ve) =  2 \cos ( 2 \theta - 2 w ) \cos^4 \left( \Frac{\ve}{2}
\right) \quad \quad \quad \quad \quad  &  \crm 
& + 2 \cos ( 2 \theta + 2 w ) \sin^4 \left( \Frac{\ve}{2} \right) + \cos (2
\theta) \sin^2 \ve \ , & \llabel{030123c}
\end{eqnarray}  
and
\be 
E_d = \frac{C - \frac{1}{2} (A + B)}{C} = \frac{ k_f R^5 }{3 G C}
\omega^2 + \delta  E_d  \ , \llabel{050109a}
\ee 
where $ E_d $ is the dynamical ellipticity, and $ k_f $ is the fluid Love number
(pertaining to a perfectly fluid body with the same mass distribution as the
actual planet).  
The first part of $E_d$ (Eq.\,\ref{050109a})
corresponds to the flattening in hydrostatic equilibrium \citep{Lambeck_1980},
and $ \delta E_d $ to the departure from this equilibrium.

\subsubsection{Averaged potential}

\llabel{030123a}

Since we are only interested in the study of the long-term motion, we 
will average the potential energy $\cU$ over the rotation angle $\theta$ and the
mean anomaly $ M $:
\be
\overline{\cU} = \frac{1}{4 \pi^2} \int_0^{2 \pi} \int_0^{2 \pi} \cU \, d M \, d
\theta \ . \llabel{100218a}
\ee
However, 
when the rotation frequency $ \omega \approx \dot \theta $ and the mean motion $ n = \dot M $
are close to resonance ($ \omega \approx p n $, for a semi-integer value $p$),
the terms with argument $ 2(\theta - p M) $ vary slowly and must be retained in
the expansions  \citep[e.g.][]{Murray_Dermott_1999}
\be 
\frac{\cos ( 2 \theta )}{r_{}^3} = \frac{1}{a^3} \sum^{+\infty}_{p=-\infty} G
( p , e ) \cos 2 (\theta - p M) \ , \llabel{061120ga}
\ee
and  
\be 
\frac{\cos ( 2 \theta - 2 w )}{r_{}^3} = \frac{1}{a^3} \sum^{+\infty}_{p=-\infty} H
( p , e ) \cos 2 (\theta - p M) \ , \llabel{061120gb}
\ee  
where  $ a $ and $ e $ are the semi-major axis and the eccentricity of the
planet's orbit, respectively. 
The functions $ G (p, e) $ and $ H (p, e) $ can be expressed in power
series in $ e $ (Table~\ref{TAB1}).
The averaged non-constant part of the potential $ \overline \cU $ becomes:
\begin{eqnarray}
\Frac{\overline \cU}{C} & = & 
- \alpha \Frac{\omega \, x^2}{2} - \Frac{\beta}{4} \left[ (1-x^2) \, G (p,e) \cos 2 (\theta
- p M) \phantom{\Frac{.}{.}} \right. \crm
&  & + \Frac{(1+x)^2}{2} \, H (p,e) \cos 2 (\theta - p M - \phi) \crm
&  & + \left. \Frac{(1-x)^2}{2} \, H (-p,e) \cos 2 (\theta - p M + \phi)
\right] \ , \llabel{030804a} 
\end{eqnarray}
where $ x = X / L = \cos \ve $, $ \phi = \varpi + \psi $,
\be 
\alpha = \Frac{3 G m_\star}{2 a^3 (1-e^2)^{3/2}} \Frac{E_d}{\omega} \approx
\Frac{3}{2} \Frac{n^2}{\omega} (1-e^2)^{-3/2} E_d \llabel{061120a}
\ee
is the ``precession constant'' and
\be 
\beta = \Frac{3 G m_\star}{2 a^3} \Frac{B - A}{C} \approx \Frac{3}{2}
n^2 \Frac{B - A}{C} \ . \llabel{030123g}
\ee

\begin{deluxetable}{l | r r r r r c | r r r r r}
\tabletypesize{\small}
\tablecaption{Coefficients of $ G ( p , e) $ and $ H ( p , e) $ to  $ e^4 $.
 \llabel{TAB1} }
\tablewidth{0pt}
\tablehead{
$ p $ & \multicolumn{5}{c}{$ G ( p , e ) $} & $\quad$ &
\multicolumn{5}{c}{$ H ( p, e ) $} \\
}
\startdata
$ \llap{$-$}1   $ & $  $ & $  $ & $ \Frac{9}{4} e^2 $ & $ + $ & $ \Frac{7}{4} e^4 $
  & $ $  & $ $ & $ $ & $  $ & $ \Frac{1}{24} e^4 $ \\ 
$ \llap{$-$}1/2 $ & $  $ & $ \Frac{3}{2} e $ & $ + $ & $ \Frac{27}{16} e^3 $ & $ $
&  & $ $ & $ $ & $  $ & $ \Frac{1}{48} e^3 $ & $ $ \\ 
$  0 $ & $ 1 $ & $ + $ & $ \Frac{3}{2} e^2 $ & $ + $ & $ \Frac{15}{8} e^4 $
&  & $ 0 $ & $  $ & $  $ & $ \phantom{\Frac{1}{1}}  $ & $ $ \\ 
$  1/2 $ & $  $ & $ \Frac{3}{2} e $ & $ + $ & $ \Frac{27}{16} e^3 $ & $ $
&  & $ - $ & $ \Frac{1}{2} e $ & $ + $ & $ \Frac{1}{16} e^3 $ & $ $ \\ 
$  1   $ & $  $ & $  $ & $ \Frac{9}{4} e^2 $ & $ + $ & $ \Frac{7}{4} e^4 $
&  & $ 1 $ & $ - $ & $ \Frac{5}{2} e^2 $ & $ + $ & $ \Frac{13}{16} e^4 $ \\ 
$  3/2 $ & $ $ & $  $ & $  $ & $ \Frac{53}{16} e^3 $ & $ $
&  & $ $ & $ \Frac{7}{2} e $ & $ - $ & $ \Frac{123}{16} e^3 $ & $ $ \\ 
$  2   $ & $ $ & $ $ & $  $ & $  $ & $ \Frac{77}{16} e^4 $
&  & $ $ & $ $ & $ \Frac{17}{2} e^2 $ & $ - $ & $ \Frac{115}{6} e^4 $ \\ 
$  5/2 $ & $ $ & $ $ & $ $ & $  $ & $ $
&  & $ $ & $ $ & $ $ & $ \Frac{845}{48} e^3 $ & $ $ \\ 
$  3   $ & $ $ & $ $ & $ $ & $ $ & $  $
&  & $ $ & $ $ & $ $ & $ $ & $ \Frac{533}{16} e^4 $ \\ 
\enddata

The exact expression of the coefficients is given by $  G ( p , e) =
\frac{1}{\pi} \int_0^\pi \left( \frac{a}{r} \right)^3 \exp(\iii \, 2 p M) \, d M
$ and $  H ( p , e) = \frac{1}{\pi} \int_0^\pi \left( \frac{a}{r} \right)^3
\exp(\iii \, 2 \nu) \exp(\iii \, 2 p M) \, d M $.
\end{deluxetable}

For non-resonant motion, that is, when $ (B-A)/C \approx 0 $ (e.g. gaseous planets)
or $ | \omega | \gg p n $,
we can simplify expression (\ref{030804a}) as:
\be
\frac{\overline \cU}{C} = - \alpha \Frac{\omega \, x^2}{2} \ . \llabel{030123e}
\ee

\subsubsection{Equations of motion}

\llabel{040811x}

The Andoyer  variables ($ L $, $ \theta$) and ($ X $, $ - \psi $) are 
canonically conjugated and thus \citep[e.g.][]{Goldstein_1950,Kinoshita_1977}
\be 
\frac{d L}{d t} = -\frac{\partial \overline \cU}{\partial \theta} \ ,
\quad \frac{d X}{d t} = \frac{\partial \overline \cU}{\partial \psi} \ ,
\quad \frac{d \psi}{d t} = - \frac{\partial \overline \cU}{\partial X} \ .
\llabel{021014c}
\ee
Andoyer's variables do not give a clear view of the spin variations, despite
their practical use. 
Since $ \omega = L/C$ and $ \cos \ve = x = X / L $ the spin variations can be
obtained as:  
\be
\frac{d \omega}{d t} = - \frac{\partial }{\partial \theta} 
\left( \frac{\overline \cU}{C} \right) \ \mathrm{,} \quad  
\frac{d \psi}{d t} = - \frac{1}{\omega} \frac{\partial}{\partial x} 
\left( \frac{\overline \cU}{C} \right)\ , \llabel{091109a} 
\ee
and
\begin{eqnarray} 
\frac{d x}{d t} &=& - \frac{1}{L} \left(
\frac{X}{L} \frac{d L}{d t} - \frac{d X}{d t} \right) \crm
&=&  \frac{1}{\omega}
\left[ x \frac{\partial}{\partial \theta} 
+ \frac{\partial}{\partial \psi} \right] \left( \frac{\overline \cU}{C} \right) 
\llabel{050602a} \ .
\end{eqnarray}

For non-resonant motion, we get from equation (\ref{030123e}):
\be 
\frac{d \omega}{d t} = \frac{d x}{d t} = 0 \quad \mathrm{and}
\quad \frac{d \psi}{d t} = \alpha \, x \ . \llabel{050214b}
\ee
The spin motion reduces to the precession of the spin vector about the
normal to the orbital plane with rate $ \alpha \, x $.

\subsection{Gravitational tides}

\llabel{030214b}

Gravitational tides arise from differential and inelastic deformations of the
planet due to the gravitational effect of a perturbing body (that can be the
central star or a satellite). 
Tidal contributions to the planet evolution are based on a very
general formulation of the tidal potential, initiated by George H. Darwin
(1880). The attraction of a body with mass $ m_{\star} $
at a distance $ r_{} $ from the center of mass of the planet can be expressed
as the gradient of a scalar potential $\cV'$, which is a sum
of Legendre polynomials \citep[e.g.][]{Kaula_1964,Efroimsky_Williams_2009}: 
\be 
\cV' = \sum_{l=2}^{\infty} \cV_l' = - \frac{G m_{\star}}{r_{}}
\sum_{l=2}^{\infty} \left(  \frac{r'}{r_{}} \right)^l P_l (\cos S) \ ,
\llabel{090419b}
\ee 
where $ r' $ is the radial distance from the planet's center, and $ S $ the
angle between $\vv{r}$ and $\vv{r}'$. 
The distortion of the planet by the potential $ \cV' $ gives rise to a tidal
potential, 
\be 
\cV_g = \sum_{l=2}^{\infty} (\cV_g)_l \ , \llabel{090419c}
\ee
where $ (\cV_g)_l = k_l \cV_l' $ at the planet's surface and $ k_l $
is the Love number for potential (Fig.\,\ref{F020423a}). 
Typically, $ k_2 \sim 0.25 $ for Earth-like
planets, and $ k_2 \sim 0.40 $ for giant planets \citep{Yoder_1995}.  
Since the tidal potential $ (\cV_g)_l $ is an $ l^\mathrm{th} $ degree harmonic,
it is a solution of a Dirichlet problem, and exterior to the planet it must be
proportional to $ r^{-l-1} $ \citep[e.g.][]{Abramowitz_Stegun_1972,
Lambeck_1980}. 
Furthermore, as upon the surface $ r' = R \ll r_{} $, we can retain in
expression (\ref{090419c}) only the first term, $ l = 2 $:
\be 
\cV_g = - k_2 \frac{G m_{\star}}{R} \left( \frac{R}{r_{}} \right)^3
\left( \frac{R}{r'} \right)^3 P_2 (\cos S) \ . \llabel{021010a}
\ee 

\begin{figure*}[t]
 \epsscale{2.}
\plotone{\figpath 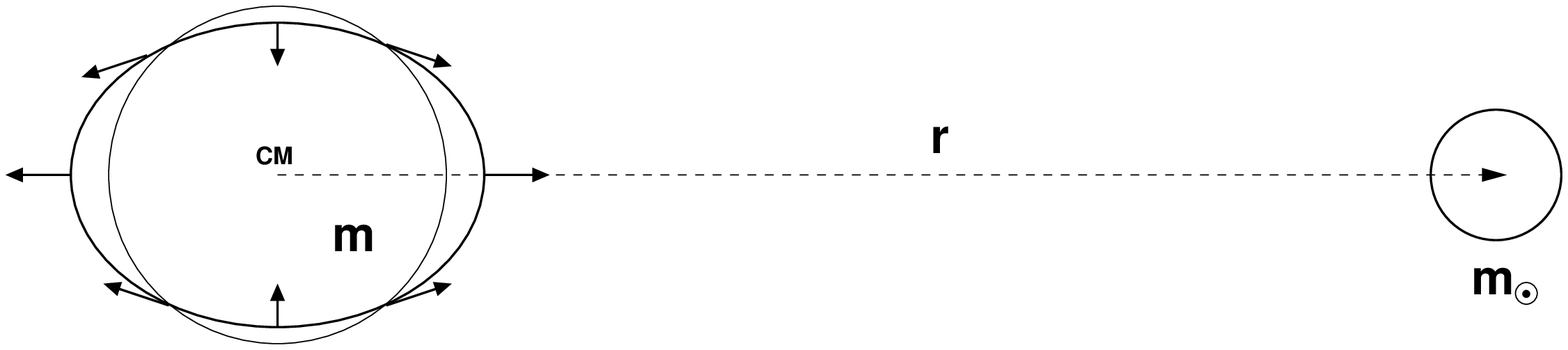}
   \caption{\small Gravitational tides. The difference between the gravitational
   force exerted by the mass $ m $ on a point of the surface and the center of
   mass is schematized by the arrows. The planet will deform following the
   equipotential of all present forces. \llabel{F020423a}}
 \end{figure*}

In general, imperfect elasticity will cause the phase angle of $ \cV_g $ to lag
behind that of $ \cV' $ \citep{Kaula_1964} by an angle $ \delta_g (\sigma) $ such
that:  
\be 
2 \delta_g (\sigma) = \sigma \Delta t_g (\sigma) \ , \llabel{090419d}
\ee
$ \Delta t_g (\sigma) $ being the time lag associated to the tidal frequency $
\sigma $ (a linear combination of the inertial rotation rate $ \omega
$ and the mean orbital motion $ n $) (Fig.\,\ref{F020423b}). 

\begin{figure*}[t]
 \epsscale{2.}
\plotone{\figpath 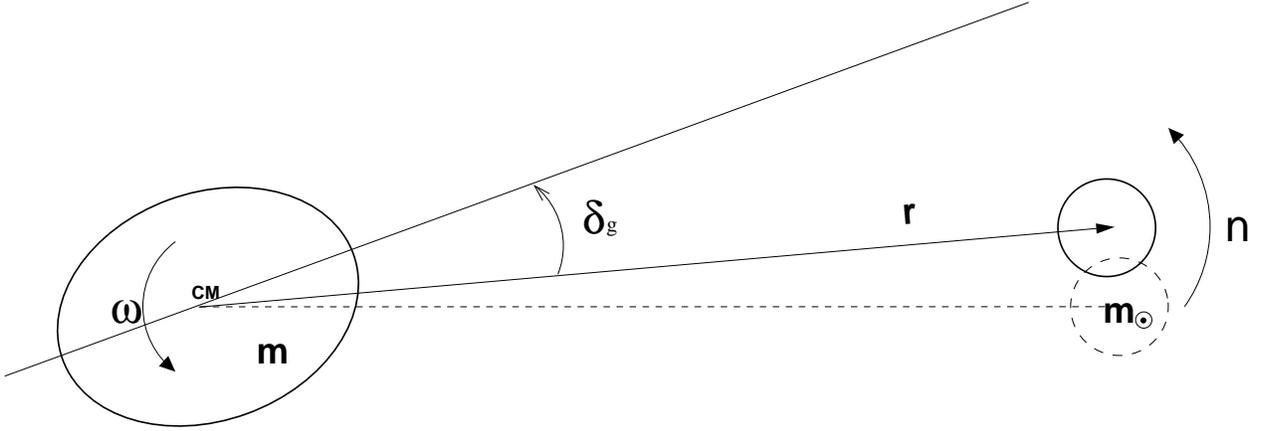}
   \caption{\small Phase lag for gravitational tides. The tidal deformation
   takes a delay time $ \Delta t_g $ to attain the equilibrium.
   During the time $ \Delta t_g $, the planet turns by an angle $ \omega \Delta t_g $ and the
   star by $ n \Delta t_g $. For $ \ve = 0 $, the bulge phase lag
   is given by $ \delta_g \approx (\omega - n) \Delta t_g $. \llabel{F020423b}}
 \end{figure*}

\subsubsection{Equations of motion}

\llabel{021205a}

Expressing the tidal potential given by expression (\ref{021010a}) in
terms of Andoyer angles $( \theta, \psi )$, we can obtain the contribution
to the spin evolution from expressions (\ref{021014c}) using $ \cU_g = m' \cV_g $
at the place of $ \overline \cU $, where $ m' $ is the mass of the interacting body.
As we are interested here in the study of the secular evolution of the spin,
we also average $ \cU_g $ over the periods of mean anomaly and longitude of
the periapse of the orbit. 
When the interacting body is the same as the perturbing body $( m' = m_\star )$,
we obtain:
\be
   \Frac{d \omega}{d t} = - \Frac{G m_{\star}^2 R^5}{C a^6} \sum_{\sigma}
                     b_g (\sigma) \Omega_\sigma^g (x, e) \ ,
 \llabel{021010c1} 
\ee
\be                   
   \Frac{d \ve}{d t} = - \Frac{G m_{\star}^2 R^5}{C a^6}
          \Frac{\sin \ve}{\omega} \sum_{\sigma} b_g (\sigma) \XX_\sigma^g (x, e) \ ,
 \llabel{021010c2} 
\ee
where the coefficients $ \Omega_\sigma^g (x, e) $ and $ \XX_\sigma^g (x, e) $
are polynomials in the eccentricity \citep{Kaula_1964}.
When the eccentricity is small, we can neglect the terms in $ e^2 $, and we have:
\be
  \begin{array}{r c l}
   \sum_\sigma b_\tau(\sigma) \Omega_\sigma^\tau 
    & = & b_\tau(\omega) \frac{3}{4}  x^2 \left( 1-x^2 \right) \\ 
    & + & b_\tau(\omega-2n) \frac{3}{16} \left(1+ x \right)^2 \left(1-x^2 \right)  \\ 
    & + & b_\tau(\omega+2n) \frac{3}{16} \left(1- x \right)^2 \left(1-x^2 \right)  \\
    & + & b_\tau(2\omega) \frac{3}{8} \left(1-x^2 \right)^2 \\ 
    & + & b_\tau(2\omega-2n) \frac{3}{32} \left(1+ x\right)^4 \\
    & + & b_\tau(2\omega+2n) \frac{3}{32} \left(1- x\right)^4 \ ,
   \end{array} 
 \llabel{V8a} 
\ee
and
\be
  \begin{array}{r c l}
   \sum_\sigma b_\tau(\sigma) \XX_\sigma^\tau 
    & = & b_\tau(2n) \frac{9}{16} \left( 1-x^2 \right) \\
    & + & b_\tau(\omega) \frac{3}{4} x^3 \\
    & - & b_\tau(\omega - 2 n) \frac{3}{16} (1+ x)^2 (2 - x) \\
    & + & b_\tau(\omega + 2 n) \frac{3}{16} (1- x)^2 (2 + x) \\
    & + & b_\tau(2 \omega) \frac{3}{8} x \left( 1-x^2 \right)  \\
    & - & b_\tau(2 \omega - 2 n) \frac{3}{32} (1+ x )^3  \\
    & + & b_\tau(2 \omega + 2 n) \frac{3}{32} (1- x )^3  \ .
   \end{array} 
 \llabel{V31} 
\ee

The coefficients $ b_\tau (\sigma) $ are related to the dissipation of the mechanical energy
of tides in the planet's interior,
responsible for the time delay $ \Delta t_g (\sigma) $ between the position of
``maximal tide'' and the sub-stellar point.  
They are related to the phase lag $ \delta_g (\sigma) $ as:
\be  
b_g (\sigma) = k_2 \sin 2 \delta_g (\sigma) = k_2 \sin
\left( \sigma \Delta t_g (\sigma) \right) \ , \llabel{021010e}
\ee 
where $\tau \equiv g $ for gravitational tides.
Dissipation equations (\ref{021010c1}) and (\ref{021010c2}) must be invariant under the change $
(\omega, x) $ by $ (- \omega, -x) $ which imposes that $ b (\sigma) = -
b (- \sigma) $, that is, $ b (\sigma) $ is an odd function of $\sigma$.
Although mathematically equivalent, the couples $ (\omega, x) $ and $ (-
\omega, -x) $ correspond to two different physical situations
\citep{Correia_Laskar_2001}.

The tidal potential given by expression (\ref{021010a})
can also be directly used to compute the orbital evolution due to tides.
Indeed, it can be seen as a perturbation of the gravitational potential
(Eq.\,\ref{090418a}), and the contributions to the orbit are computed using
Lagrange Planetary equations \citep[e.g.][]{Brouwer_Clemence_1961,Kaula_1964}:
 \begin{eqnarray} 
 \frac{d a}{d t} & = & \frac{2}{\mu n a} \frac{\partial U}
{\partial M} \ , \llabel{010705b} \\
\frac{d e}{d t} & = & \frac{\sqrt{1- e^2}}{\mu n a^2 e} \left[ \sqrt{1-e^2}
\frac{\partial U}{\partial M} - \frac{\partial U}{\partial \varpi} \right] \ ,
\llabel{010705c} 
\end{eqnarray} 
where $ \mu = m m_\star / (m + m_\star) \approx m $ is the reduced mass.

We then find for the orbital evolution of the planet:
\be
   \Frac{d a}{d t} = - \frac{6 G m_\star^2 R^5}{\mu n a^7}  \sum_{\sigma}
                     b_g (\sigma) A_\sigma^g (x, e) \ ,
 \llabel{100208a} 
\ee
\be                   
   \Frac{d e}{d t} = - e \, \frac{3 G m_\star^2 R^5}{\mu n a^8} 
           \sum_{\sigma} b_g (\sigma) E_\sigma^g (x, e) \ ,
 \llabel{100208b} 
\ee
where the coefficients $ A_\sigma^g (x, e) $ and $ E_\sigma^g (x, e) $
are again polynomials in the eccentricity.
When the eccentricity is small, we can neglect the terms in $ e^2 $, and we have:
\be
  \begin{array}{r c l}
   \sum_\sigma b_\tau(\sigma) A_\sigma^\tau 
    & = & b_\tau(2 n) \frac{9}{16} (1-x^2)^2   \\
    & - & b_\tau(\omega-2n) \frac{3}{8} (1-x^2)(1+ x)^2  \\
    & + & b_\tau(\omega+2n)  \frac{3}{8} (1-x^2)(1- x)^2 \\ 
    & - & b_\tau(2\omega-2n) \frac{3}{32} (1+ x)^4 \\ 
    & + & b_\tau(2\omega+2n) \frac{3}{32} (1- x)^4  \ ,
   \end{array} 
 \llabel{100208c} 
\ee
and
\be
  \begin{array}{r c l}
   \sum_\sigma b_\tau(\sigma) E_\sigma^\tau 
& = & b_\tau (n) \frac{9}{128} (5 x^2-1)(7 x^2-3) \\
& - & b_\tau (2 n) \frac{9}{32} (1-x^2)^2  \\
& + & b_\tau (3 n) \frac{441}{128} (1-x^2)^2  \\
& - & b_\tau (\omega - n) \frac{3}{64} (5 x-1)(7 x+1)(1-x^2) \\
& + & b_\tau (\omega + n) \frac{3}{64} (5 x+1)(7 x-1)(1-x^2) \\ 
& + & b_\tau (\omega-2n) \frac{3}{16} (1-x^2)(1+x)^2 \\ 
& - & b_\tau (\omega+2n) \frac{3}{16} (1-x^2)(1-x)^2 \\ 
& - & b_\tau (\omega-3n) \frac{3}{64} (1-x^2)(1+x)^2  \\
& + & b_\tau (\omega+3n) \frac{3}{64} (1-x^2)(1-x)^2 \\   
& - & b_\tau (2\omega-n) \frac{3}{256} (5 x-7)(7 x-5)(1+x)^2 \\ 
& + & b_\tau (2\omega+n) \frac{3}{256} (5 x+7)(7 x+5)(1-x)^2 \\ 
& + & b_\tau (2\omega-2n) \frac{3}{64} (1+x)^4 \\ 
& - & b_\tau (2\omega+2n) \frac{3}{64} (1-x)^4 \\ 
& - & b_\tau (2\omega-3n) \frac{147}{256} (1+x)^4 \\
& + & b_\tau (2\omega+3n) \frac{147}{256} (1-x)^4  \ .
   \end{array} 
 \llabel{100208d} 
\ee

\subsubsection{Dissipation models}

\llabel{021024j}

The dissipation of the mechanical energy of tides in the planet's interior is
responsible for the phase lags $ \delta (\sigma) $.
A commonly used dimensionless measure of tidal damping is the quality factor $
Q $ \citep{Munk_MacDonald_1960}, defined as the inverse of the ``specific''
dissipation and related to the phase lags by
\be 
Q (\sigma) = \frac{2 \pi E}{\Delta E} = \cot 2 \delta (\sigma) \ ,
\llabel{090419e}
\ee
where $ E $ is the total tidal energy stored in the planet, and $ \Delta
E $ the energy dissipated per cycle. We can rewrite expression (\ref{021010e})
as: 
\be 
b_g (\sigma) = \frac{ k_2 \, \mathrm{sign}(\sigma)}{\sqrt{Q^2 (\sigma) + 1}}
\approx \mathrm{sign}(\sigma) \frac{k_2}{Q (\sigma)} \ .
\ee 
The present $ Q $ value for the planets in the Solar system
can be estimated from orbital measurements,
but as rheology of the planets is  badly known,
the exact dependence of $ b_\tau (\sigma)$ on the tidal frequency $ \sigma $
is unknown.
Many different authors have studied the problem and several models for $ b_\tau
(\sigma)$  have been developed so far, from the simplest ones to the more
complex \citep[for a review see][]{Efroimsky_Williams_2009}.
The huge problem in validating one model better than the others is the difficulty
to compare the theoretical results with the observations, as the effect
of tides are very small and can only be detected efficiently after long periods
of time.
Therefore, here we will only describe a few simplified models that are commonly
used:

\paragraph{The visco-elastic model}

\llabel{021120y}

\citet{Darwin_1908} assumed that the planet behaves like a Maxwell
solid, that is, the planet responds to stresses
like a massless, damped harmonic oscillator. It is characterized by a rigidity
(or shear modulus) $ \mu_e $ and by a viscosity $ \upsilon_e $. A Maxwell solid
behaves like an elastic solid over short time scales, but flows like a fluid
over long periods of time. This behavior is also known as elasticoviscosity. 
For a constant density $ \rho $, we have:
\be
b_g (\sigma) = k_f \frac{\tau_b - \tau_a}{1 + (\tau_b \, \sigma)^2} \sigma \ ,
\llabel{090419f}
\ee
where $ k_f $ is the fluid Love number (Eq.\,\ref{050109a}).
$ \tau_a = \upsilon_e / \mu_e $ and $ \tau_b = \tau_a ( 1 + 19 \mu_e R / 2 G m
\rho )  $ are time constants for the damping of gravitational tides.

The visco-elastic model is a realistic approximation of the planet's
deformation with the tidal frequency \citep[e.g.][]{Escribano_etal_2008}.
However, when replacing expression (\ref{090419f}) into the dynamical
equations (\ref{021010c1}) and (\ref{021010c2}) we get an infinite sum of terms,
which is not practical.
As a consequence,  simplified versions of the visco-elastic model
for specific values of the tidal frequency $ \sigma $ are often used.
For instance, when $\sigma$ is small, $ (\tau_b \, \sigma)^2$ can be neglected in
expression (\ref{090419f}) and $b_g(\sigma)$ becomes proportional to $\sigma$.

\paragraph{The viscous or linear model}

\llabel{021120a}

In the viscous model, 
it is assumed that the response time delay to the perturbation is 
independent of the tidal frequency, i.e., the position of the ``maximal
tide'' is shifted from the sub-stellar point by a constant time lag $ \Delta t_g $
 \citep{Mignard_1979,Mignard_1980}. As usually we have  $ \sigma \Delta t_g \ll 1 $,
the viscous model becomes linear:
\be 
b_g (\sigma) = k_2 \sin (\sigma \Delta t_g) \approx k_2 \, \sigma \Delta t_g \ .
\llabel{091126a}
\ee  
The viscous model is a particular case of the visco-elastic model and is
specially adapted to describe the behavior of planets in slow rotating regimes
($ \omega \sim n $).

\paragraph{The constant-$Q$ model}

\llabel{021024i}

Since for the Earth, $ Q $ changes by less than an order of magnitude between the
Chandler wobble period (about 440 days) and seismic periods of a few
seconds \citep{Munk_MacDonald_1960}, it is also common to treat the specific
dissipation as independent of frequency. Thus,
\be 
b_g (\sigma) \approx \mathrm{sign}(\sigma) k_2 / Q \ . \llabel{090407a}
\ee
The constant-$Q$ model can be used for periods of time where the tidal frequency does
not change much, as is the case for fast rotating planets. However, for
long-term evolutions and slow rotating planets, the constant-$Q$ model is 
not appropriate as it gives rise to discontinuities for $ \sigma = 0 $.

\subsubsection{Consequences to the spin}

\llabel{011012a}

Although both linear and constant models have some limitations, for simplicity
reasons they are the most widely used in literature. 
The linear model has nevertheless an important advantage over the constant model:
it is appropriate do describe the behavior of the planet near the equilibrium
positions, since the linear model closely follows the realistic visco-elastic
model for slow rotation rates.
The equations of motion can also be expressed in an elegant way, so we will
adopt the viscous model for the remaining of this Chapter, without loss of
generality concerning the main consequences of tidal effects.

Using the approximation (\ref{091126a}) in expressions (\ref{021010c1}) and
(\ref{021010c2}), we simplify the spin equations as
\citep[][Appendix B]{Correia_Laskar_2010}: 
\be
\dot \omega = - \frac{\Kn}{C} 
\left( f_1(e) \Frac{1 + \cos^2 \ve}{2} \Frac{\omega}{n} -
f_2(e) \cos \ve \right) \ , \llabel{090515a}
\ee
and
\be
\dot \ve \approx \frac{\Kn}{C \omega} \sin \ve
\left( f_1(e) \cos \ve \frac{\omega}{2 n} - f_2(e) \right) 
\ , \llabel{090520d}
\ee
where
\be
f_1(e) = \Frac{1 + 3 e^2 + 3 e^4 / 8}{(1 - e^2)^{9/2}} \ , \llabel{030218a}
\ee
\be
f_2 (e) = \Frac{1 + 15 e^2 / 2 + 45 e^4 / 8+ 5 e^6 / 16}{(1 - e^2)^{6}} \ ,
\llabel{030218b}
\ee
and
\be
K = \Delta t \frac{3 k_2 G m_\star^2 R^5}{a^6}  \ . \llabel{eq3} 
\ee
Because of the factor $ 1 / \omega $ in the magnitude of the obliquity
variations (Eq.\,\ref{090520d}), for an initial fast rotating planet the 
time-scale for the obliquity evolution will be longer than the time-scale
for the rotation rate evolution (Eq.\,\ref{090515a}). 
As a consequence, it is to be expected that the rotation rate reaches an
equilibrium value earlier than the obliquity.
For a given obliquity and eccentricity, the equilibrium rotation rate, obtained
when $ \dot \omega = 0 $, is then attained for (Fig.\,\ref{FigC}):
\be
\frac{\omega_e}{n} = \frac{f_2 (e)}{f_1 (e)} \, \frac{2 \cos
\ve}{1 + \cos^2 \ve} \ , \llabel{090520a}
\ee
Replacing the previous equation in the expression for obliquity variations
(Eq.\,\ref{090520d}), we find:
\be
\dot \ve \approx - \frac{\Kn}{C \omega} f_2(e) \frac{\sin \ve}{1+\cos^2 \ve} \ .
\llabel{090520e}
\ee
We then conclude that the obliquity can only decrease by tidal effect,
since $ \dot \ve \le 0 $, and the final obliquity always tends to zero.

\begin{figure}[t]
 \epsscale{1.}
\plotone{\figpath 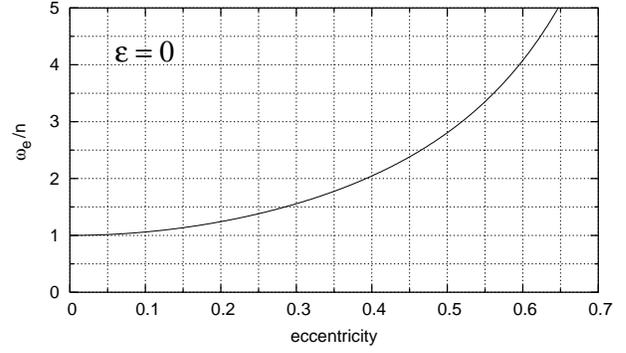}
  \caption{\small Evolution of the equilibrium rotation rate $ \omega_e / n =
  f_2(e)/f_1(e) $ with the eccentricity when $ \ve = 0^\circ $ using the viscous
  model (Eq.\,\ref{090520a}). 
  As the eccentricity increases, $ \omega_e $ also increases. The gravitational tides
  lead the planet to exact resonance when the eccentricity is respectively
  $ e_{1/1} = 0 $, $ e_{3/2} = 0.284926803 $ and $ e_{2/1} = 0.392363112 $.
   \llabel{FigC} }
 \end{figure}

\subsubsection{Consequences to the orbit}

As for the spin, the semi-major axis and the eccentricity evolution can be
obtained using the approximation (\ref{091126a}) in expressions (\ref{100208c})
and (\ref{100208d}), respectively \citep{Correia_2009}: 

\be
\dot a = \frac{2 K}{\mu a} \, \left( f_2(e) \cos \ve \frac{\omega}{n} -
f_3(e) \right) \ , \llabel{090515b} 
\ee
and\be
\dot e = \frac{9 K}{\mu a^2} \left( \frac{11}{18} f_4(e) \cos
\ve \frac{\omega}{n} - f_5(e) \right) e \ , \llabel{090515c} 
\ee
where 
\be
f_3(e) = \frac{1 + 31e^2/2 + 255e^4/8 + 185e^6/16 + 25e^8/64}{(1-e^2)^{15/2}} 
\ , \llabel{090514p}
\ee
\be
f_4(e) = \frac{1 + 3e^2/2 + e^4/8}{(1-e^2)^5} \ , \llabel{090515d}
\ee
\be
f_5(e) = \frac{1 + 15e^2/4 + 15e^4/8 + 5e^6/64}{(1-e^2)^{13/2}} \ . \llabel{090515e}
\ee

The ratio between orbital and spin evolution time-scales is roughly given by 
$ C / (\mu a^2) \ll 1 $, meaning that the spin achieves an equilibrium position
much faster than the orbit.
Replacing the equilibrium rotation rate (Eq.\,\ref{090520a}) with $ \ve = 0 $
(for simplicity) in equations (\ref{090515b}) and (\ref{090515c}), gives:
\be
\dot a = - \frac{7 K}{\mu a} \, f_6(e) e^2 \ , \llabel{090522a} 
\ee
\be
\dot e = - \frac{7 K}{2 \mu a^2} f_6(e) (1-e^2) e \ , \llabel{090522b} 
\ee
where 
$ 
f_6 (e) = (1 + 45e^2/14 + 8e^4 + 685e^6/224 + 255e^8/448 + 25e^{10}/1792)
(1-e^2)^{-15/2} / (1 + 3e^2 + 3e^4/8)  \ . \llabel{090527a}
$
Thus, we always have $ \dot a \le 0 $ and $ \dot e \le 0 $, and the
final eccentricity is zero. 
Another consequence is that the quantity $ a (1 - e^2) $ is conserved
(Eq.\,\ref{100210z}).
The final equilibrium semi-major axis is then given by
\be
a_f = a (1 - e^2) \ , \llabel{090522c}
\ee
which is a natural consequence of the orbital angular momentum conservation
(since the rotational angular momentum of the planet is much smaller).
Notice, however, that once the equilibrium semi-major axis $ a_f $ is attained, the tidal
effects on the star cannot be neglected, and they govern the future evolution of
the planet's orbit.

\subsection{Thermal atmospheric tides}

\llabel{RAT}

The differential absorption of the Solar heat by the planet's atmosphere gives 
rise to local variations of temperature and consequently to pressure gradients. 
The mass of the atmosphere is then permanently redistributed, adjusting for an
equilibrium position.
More precisely, the particles of the atmosphere move from the high temperature
zone (at the sub-stellar point) to the low temperature areas.
Indeed, observations on Earth show that the pressure redistribution is
essentially a superposition of two pressure waves 
\citep[see][]{Chapman_Lindzen_1970}:
a daily (or diurnal) tide of small amplitude (the pressure is minimal at the
sub-stellar point and maximal at the antipode) and a strong half-daily (semi-diurnal) tide (the
pressure is minimal at the sub-stellar point and at the antipode)  (Fig.\,\ref{010430.f1}).

The gravitational potential generated by all of the particles in the atmosphere 
at a generic point of the space $ \vv{r} $ is given by:  
\be 
\cV_a = - G \int\limits_{({\cal M})}^{} \frac{d {\cal M}}{\md \vv{r} - \vv{r}'
\md} \ , \llabel{A1} 
\ee 
where $ \vv{r}' = (r', \theta', \varphi') $ is the position of the atmosphere
mass element $ d {\cal M} $ with density $ \rho_a (\vv{r}') $ and
\be 
d {\cal M} = \rho_a (\vv{r}') {r'}^2 \sin \theta' \, d r' \, d \theta' \, d
\varphi' \ . \llabel{A0} 
\ee 

Assuming that the radius of the planet is constant
and that the height of the atmosphere can be neglected,
we  approximate  expression (\ref{A0}) as:
\be 
d {\cal M} = \frac{{R_{}}^2}{g} p_s (\theta', \varphi', t) \sin \theta' \, d
\theta' \, d \varphi' \ , \llabel{'msga'}
\ee
where $ g $ is the mean surface gravity acceleration, and 
$ p_s $   the surface pressure, which depends on the stellar insolation.
Thus, $ p_s $ depends on $ S $, the angle between the direction of the Sun and the
normal to the surface: 
\be 
p_s (\theta', \varphi') = p_s(S) = \sum_{l=0}^{+\infty}
\tilde{p}_l \, P_l(\cos S) \ , \llabel{A2} 
\ee
where $ P_l $ are the Legendre polynomials of order $ l $ and 
$ \tilde{p}_l $ its coefficients. 
Developing also $ \md \vv{r} - \vv{r}' \md^{-1} $ in   
Legendre polynomials we rewrite expression (\ref{A1}) as:
\be 
\cV_a = - \frac{1}{\bar{\rho}} \sum_{l=0}^{+\infty} \frac{3}{2 l + 1} \,
\tilde{p}_l \left( \frac{R}{r} \right)^{l+1} P_l(\cos S) \ , \llabel{A3} 
\ee
where $ \bar{\rho} $ is the mean density of the planet.
Since we are only interested in pressure oscillations, we must subtract the term
of constant pressure ($ l=0 $) in order to obtain the tidal potential.
We also eliminate the diurnal terms ($ l=1 $) because they correspond to a
displacement of the center of mass of the atmosphere bulge which has no
dynamical implications.  
Thus, since we usually have $ r \gg R $, retaining only the semi-diurnal terms
($ l = 2 $), we write:
\be 
\cV_a = - \frac{3}{5} \frac{\tilde{p}_2}{\bar{\rho}} \left( 
\frac{R_{}}{r} \right)^3 P_2(\cos S) \ . \llabel{A5} 
\ee

\begin{figure*}[t]
 \epsscale{1.9}
\plotone{\figpath 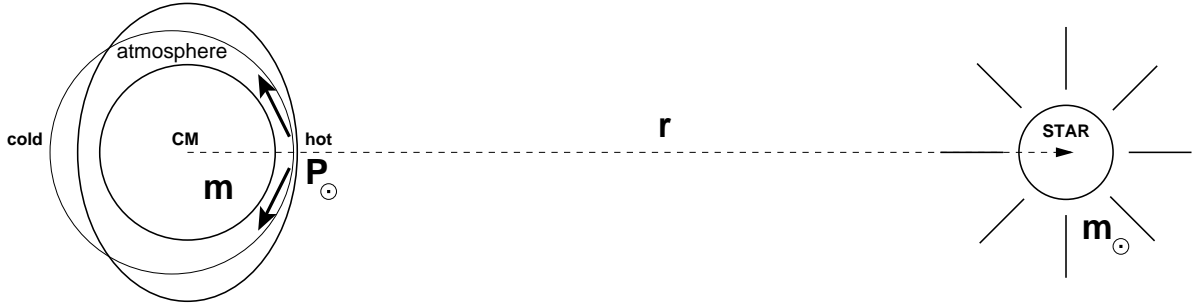}
   \caption{\small Thermal atmospheric tides. The atmosphere's heating
   decreases with the distance to the sub-stellar point $ P_\odot $. The
   atmospheric mass redistribution is essentially decomposed in a weak
   daily tide (round shape) and in a strong
   half-daily tide (oval shape). \llabel{010430.f1}}
 \end{figure*}

\subsubsection{Equations of motion}

Using the same methodology of previous sections, the contributions of thermal
atmospheric tides to the spin evolution are obtained 
from expressions (\ref{021014c}) using $ \cU_a = m_\star \cV_a $
at the place of $ \overline \cU $:
\be
   \Frac{d \omega}{d t} = - \Frac{3 m_\star {R_{}}^3}{5 C \bar{\rho} a^3} 
                     \sum_{\sigma} b_a (\sigma) \Omega_\sigma^a (x, e) \ ,
 \llabel{A6a} 
\ee
\be                   
   \Frac{d \ve}{d t} = - \Frac{3 m_\star {R_{}}^3}{5 C \bar{\rho} a^3} 
          \Frac{\sin \ve}{\omega} \sum_{\sigma} b_a (\sigma) \XX_\sigma^a (x, e) \ ,
 \llabel{A6b} 
\ee
where the terms $ \Omega_\sigma^a (x, e)$ and $ \XX_\sigma^a (x,e) $ are
also polynomials in the eccentricity, but different from their
analogs for gravitational tides (Eqs.\,\ref{021010c1} and \ref{021010c2}). 
Nevertheless, when neglecting the terms in $ e^2 $, they become equal and are
given by expressions (\ref{V8a}) and (\ref{V31}), respectively (with $\tau = a $).

For thermal atmospheric tides there is also a delay before the response of the
atmosphere to the excitation (Fig.\,\ref{010430.f2}).
We name the time delay $ \Delta t_a (\sigma) $ and the corresponding phase angle $
\delta_a (\sigma) $ (Eq.\,\ref{090419d}).
The dissipation factor $ b_a (\sigma) $ is here given by:
\be 
b_a (\sigma) = \tilde{p}_2 (\sigma) \sin 2 \delta_a (\sigma) = 
\tilde{p}_2 (\sigma) \sin( \sigma \Delta t_a (\sigma) ) \ . \llabel{A9} 
\ee 

\citet{Siebert_1961} and \citet{Chapman_Lindzen_1970} have shown that when 
\be 
\md \tilde{p}_2 (\sigma) \md \ll \tilde{p}_0 \ , \llabel{010805a}
\ee 
the amplitudes of the pressure variations on the ground are given by:
\be 
\tilde{p}_2 (\sigma) = \iii \frac{\gamma}{\sigma} \tilde{p}_0 \left(
\nabla \cdot \vv{v}_\sigma - \frac{\gamma - 1}{\gamma} \frac{J_\sigma}{g H_0}
\right) \ , \llabel{A8} 
\ee
where $ \gamma = 7/5 $ for a perfect gas, $ \vv{v} $ is the velocity of tidal
winds, $ J_\sigma $ the amount of heat absorbed or emitted by a unit mass of air per
unit time, and $ H_0 $ is the scale height at the surface. 
We can rewrite expression (\ref{A8}) as,
\begin{eqnarray} 
\tilde{p}_2 (\sigma) & = & \frac{\gamma}{\md \sigma \md}
\tilde{p}_0 \left\md \nabla \cdot \vv{v}_\sigma - \frac{\gamma - 1}{\gamma}
\frac{J_\sigma}{g H_0} \right\md e^{\pm \iii \frac{\pi}{2}} \nonumber \\
\llabel{A10} \\ & = & \md \tilde{p}_2 (\sigma) \md e^{\pm \iii
\frac{\pi}{2}} \ , \nonumber 
\end{eqnarray} 
where the factor $ e^{\pm \iii \frac{\pi}{2}} $ can be seen as a supplementary
phase lag of $ \pm \pi / 2 $:
\begin{eqnarray} 
b_a (\sigma) & = & \md \tilde{p}_2 (\sigma) \md 
 \sin 2 \left( \delta_a (\sigma) \pm \frac{\pi}{2} \right) \crm 
 & = & - \md \tilde{p}_2 (\sigma) \md \sin 2 \delta_a (\sigma) \ . \llabel{VV9} 
\end{eqnarray} 
The minus sign above causes pressure variations to lead the Sun whenever $
\delta_a (\sigma) < \pi / 2 $ \citep{Chapman_Lindzen_1970,
Dobrovolskis_Ingersoll_1980} (Fig.\,\ref{010430.f2}). 

\begin{figure*}[t]
 \epsscale{1.9}
\plotone{\figpath 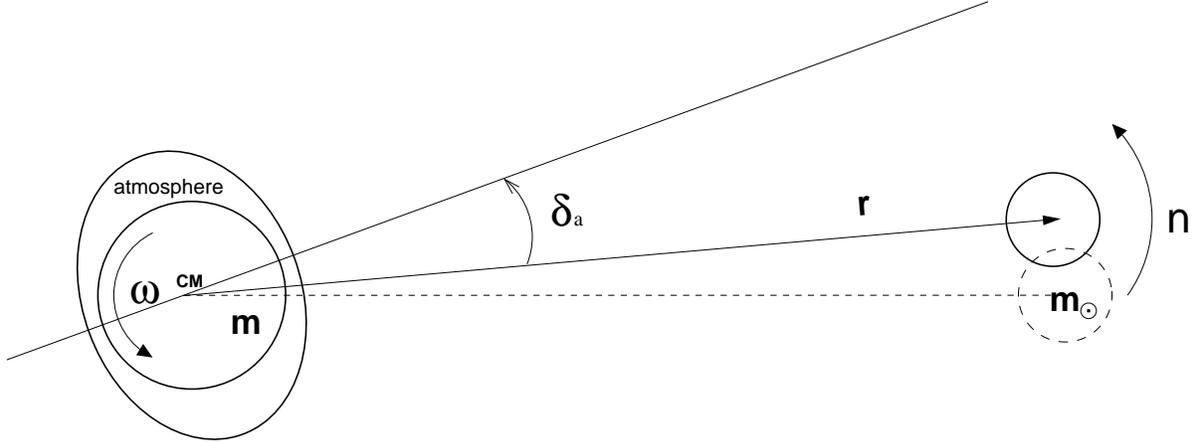}
   \caption{\small Phase lag for thermal atmospheric tides. 
   During the time $ \Delta t_a $ the planet turns by an angle $ \omega \Delta
   t_a $ and the star by $ n \Delta t_a $. For $ \ve = 0 $, the bulge phase lag
   is given by $ \delta_a \approx (\omega - n) \Delta t_a $. \llabel{010430.f2}}
 \end{figure*}

\subsubsection{Dissipation models}

Unfortunately, our knowledge of the atmosphere response to thermal excitation is
still very incomplete.
As for the gravitational tides, models are developed to deal with the unknown part.
\citet{Dobrovolskis_Ingersoll_1980} adopted a model called ``{\it
heating at the ground}'', where they suppose that all the stellar flux absorbed by
the ground  $ F_s $ is immediately deposited  in a thin layer of atmosphere at
the surface.  
The heating distributing may then be written as a delta-function just
above the ground:
\be 
J ( r ) = \frac{g}{\tilde{p}_0} F_s \delta ( r - 0^+ ) \ . 
\ee

Neglecting $ \vv{v} $ over the thin heated layer, expression (\ref{A8})
simplifies: 
\be 
\md \tilde{p}_2 (\sigma) \md = \frac{5}{16} \frac{\gamma - 1} { \md \sigma
\md }  \frac{F_s}{H_0} = \frac{5}{16} \frac{\gamma}{\md \sigma \md}  \frac{g
F_s}{c_p \bar{T}_s} \ , \llabel{A11} 
\ee 
where the factor $ 5 / 16 $ represents the second-degree harmonic component
of the insolation contribution \citep{Dobrovolskis_Ingersoll_1980} and $ c_p $ is
the specific heat at constant pressure. 

Nevertheless, according to expression (\ref{A11}), if $ \sigma = 0 $, the
amplitude of the pressure variations $ \tilde{p}_2 (\sigma) $ becomes infinite. 
The amplitude cannot grow infinitely, as for a tidal frequency
equal to zero, a steady distribution is attained, and thus $ \tilde{p}_2 (0) = 0
$.
Indeed, expression (\ref{A8}) is not valid when $ \sigma \approx 0 $ because the
condition (\ref{010805a}) is no longer verified.
Using the typically accepted values for the Venusian atmosphere,
$ c_p \approx 1000 \, \mathrm{K} \, \mathrm{kg}^{-1} \mathrm{s}^{-1} $, $
\bar{T}_s \approx 730 \, \mathrm{K} $ and $ F_s \approx 100 \, \mathrm{Wm}^{-2} $
\citep{Avduevskii_etal_1976} we compute:
\be 
\md \tilde{p}_2 (\sigma) \md \approx 10^{-4} \, \tilde{p}_0 \frac{n}{\md
\sigma \md} \ , \llabel{091126b}
\ee
which means that for $ \sigma \sim n $, the ``{\it heating at the ground}'' model of 
\citet{Dobrovolskis_Ingersoll_1980}, can still be applied.
Since we are only interested in long-term behaviors we can set
$ \tilde{p}_2 (\sigma) = 0 $ whenever $ \md \sigma \md \ll n / 100 $.
Moreover, for tidal frequencies $ \sigma \sim 0 $ the dissipation lag $\sin
(\sigma \Delta t_a (\sigma)) \approx \sigma \Delta t_a (\sigma) $ also goes to zero. 
We expect that further studies about synchronous exoplanets'
atmospheres \citep[e.g.][]{Joshi_etal_1997,Arras_Socrates_2010b},
may provide a more accurate solution for the case $ \sigma \approx 0 $. 

In presence of a dense atmosphere, another kind of tides
 can arise: the atmosphere pressure upon the surface gives rise to a
deformation, a pressure bulge, that will also be affected by the stellar torque.
At the same time, the atmosphere itself exerts a torque over the planet's
bulges (gravitational and pressure bulge).
Nevertheless, we do not need to take into account additional tidal effects as
their consequences upon the dynamical equations can be neglected
\citep{Hinderer_etal_1987,Correia_Laskar_2003JGR}.

\subsection{Spin-orbit resonances}

\llabel{RSOR}

A spin-orbit resonance occurs when there is a commensurability between the
rotation rate $ \omega $ and the mean motion of the orbit $ n $ (Eqs.\,\ref{061120ga} and \ref{061120gb}). 
The synchronous rotation of the Moon is the  most common example. 
After the discovery of the 3/2 spin-orbit resonance of Mercury
\citep{Colombo_1965}, spin-orbit resonances were studied in great detail 
\citep{Colombo_Shapiro_1966,Goldreich_Peale_1966,
Counselman_Shapiro_1970,Correia_Laskar_2004,Correia_Laskar_2009,Correia_Laskar_2010}.
When resonant motion is present we cannot neglect the terms in $ \beta $ in
expression (\ref{030804a}).
Assuming for simplicity a low obliquity ($ x \approx 1 $) we obtain a
non-zero contribution for the rotation rate (Eq.\,\ref{091109a}):
\be 
\frac{d \omega}{d t} = - \beta \, H(p,e) \sin 2 \gamma  \llabel{Res1} \ , 
\ee 
where $ \gamma = \theta - p M - \phi $.
The rotation of the planet will therefore present oscillations around a mean
value.
The width of the corresponding resonance, centered at $ \omega = p n $, is:
\be 
\Delta \omega = \sqrt{2 \beta \, H(p,e)} \ . \llabel{Res1b} 
\ee  

Due to the tidal torque (Eq.\,\ref{021010c1}), here denoted by $\overline T$, the
mean rotation rate does not remain constant and may therefore cross and be
captured in a spin-orbit resonance.
\citet{Goldreich_Peale_1966} computed a simple estimation of 
the capture probability $ P_\mathrm{cap} $,  
and subsequent more detailed studies proved their expression to be
essentially correct \citep[for a review, see][]{Henrard_1993}. 
Since the tidal torques can usually be described by means of
the torques considered by \citet{Goldreich_Peale_1966}, we will adopt here the
same notations. 
Let 
\be 
\overline T = - K \left( V + \frac{\dot \gamma}{n}
\right) \llabel{Res2} \ , 
\ee  
where $ K $ and $ V $ are positive constant torques, and $ \dot \gamma = \omega
- p n $. 
The probability of capture into resonance is then given by
\citep{Goldreich_Peale_1966}: 
\be
 P_\mathrm{cap} = \Frac{2}{1 + \pi V n / \Delta \omega} \ ,
 \llabel{Res3} 
\ee
where $\Delta \omega $ is the resonance width (Eq.\,\ref{Res1b}).
In the slow rotation regime ($ \omega \sim n $), where
the spin encounters spin-orbit resonances and capture may occur,
we compute for the viscous tidal model (Eq.\,\ref{090515a}):
\be
P_\mathrm{cap} = 2 \left[ 1 + \left( p - \Frac{2 x}{1+x^2} \Frac{f_2(e)}{f_1(e)}
\right) \Frac{n \pi}{2 \Delta \omega} \right]^{-1} \ , \llabel{040819b} 
\ee

\subsection{Planetary perturbations}

Like in the Solar System, many exoplanets are not alone in their orbits, but
belong to multi-planet systems. 
Because of mutual planetary perturbations the orbital parameters of the planets
do not remain constant and undergo secular variations in time
(Chapter~10: {\it Non-Keplerian Dynamics}).
An important consequence for the spin of the planets is that the reference
orbital plane (to which the obliquity and the precession were defined) will also
present variations.
We can track the orbital plane variations by the inclination to an
inertial reference plane, $I$, and by the longitude of the line of nodes, $ \Omega $.
Under the assumption of principal axis rotation, the energy perturbation
attached to an inertial frame can be written
\citep{Kinoshita_1977,Surgy_Laskar_1997}: 
\begin{eqnarray}
\cU_{pp} & = & \left[ X (1 - \cos I) 
- L \sin \ve \sin I \cos \varphi \right] \Frac{d \Omega}{d t}  \crm  
& + & L \sin \ve \sin \varphi \Frac{d I}{d t}  \ ,
\llabel{050819a}
\end{eqnarray}
where $ \varphi = - \Omega - \psi $.

Although the Solar System motion is chaotic \citep{Laskar_1989,Laskar_1990}, 
the motion can be approximated over several million of years 
by quasi-periodic series. In particular for the orbital elements that 
are involved in the precession driving terms (Eq.\,\ref{050819a}), we have 
 \citep{Laskar_Robutel_1993}:
\be
\bigg(\Frac{d I}{dt}+\iii\,\Frac{d \Omega}{dt}\,\sin I \bigg)\,e^{\iii\Omega}=
\sum_{k} {\cal J}_k\,e^{\iii(\nu_k t+\phi_k)} \ , \llabel{050820b}
\ee
and
\be
(1 - \cos I ) \, \Frac{d \Omega}{dt} = \sum_k {\cal L}_k \cos (\nu_k t +
\varphi_k) \ , \llabel{060213a}
\ee
where $\nu_k$ are secular frequencies of the orbital motion with
amplitude ${\cal J}_k$ and phase $\phi_k$, and $\iii=\sqrt{-1}$.
We may then rewrite expression (\ref{050819a}) as
\begin{eqnarray}
\cU_{pp} & = & L \sum_k \left[ {\cal L}_k \cos (\nu_k t + \varphi_k) \, x
\phantom{\frac{}{}} \right. \crm
& & \quad \left. - {\cal J}_k \sqrt{1-x^2} \, \sin
(\nu_k t + \psi + \phi_k) \right] \ . \llabel{091115a}
\end{eqnarray}

Assuming non-resonant motion, from equations (\ref{091109a}) and
(\ref{050602a}) we get for the spin motion:
\be
\Frac{d \ve}{d t} = \sum_k {\cal J}_k \cos( \nu_k t + \psi + \phi_k) \ , 
\llabel{050820c}
\ee
and
\begin{eqnarray}
\Frac{d \psi}{d t} & = & \alpha \cos \ve 
- \sum_k {\cal L}_k \cos(\nu_k t + \varphi_k ) \crm 
& & - \cot \ve \sum_k {\cal J}_k \sin (\nu_k t + \psi + \phi_k) \ .
\llabel{060213b}
\end{eqnarray}

For planetary systems like the Solar System, the mutual inclinations remain
small \citep{Laskar_1990,Correia_etal_2010}, and 
it follows from expressions (\ref{050820b}) and (\ref{060213a}) that the
amplitudes of $ {\cal J}_k $ and $ {\cal L}_k $ are bounded respectively by
\be
{\cal J}_k \sim \nu_k \, I_{max}  \ , \quad {\cal L}_k \sim \nu_k  \, I_{max}^2 / 2 
\ . \llabel{060213c}
\ee
The term in $ {\cal L}_k $ in expression (\ref{060213b}) for the precession variations can
then be neglected for small inclinations.

From expression (\ref{050820c}) it is clear that a 
resonance can occur whenever the precession frequency is equal 
to the opposite of a secular frequency $\nu_k$ (that is, $ \dot \psi = - \nu_k $).
Retaining only the terms in $ k $, the problem becomes integrable.
We can search for the equilibrium positions by setting
the obliquity variations equal to zero ($ \dot \ve = 0 $).
It follows then from expression (\ref{050820c}) that $ \psi + \nu_k t + \phi_k =
\pm \pi / 2 $, and
replacing it in expression (\ref{060213b}) with $ \dot \psi = - \nu_k $,
we get
\be
\alpha \cos \ve \sin \ve + \nu_k \sin \ve \approx {\cal J}_k \cos \ve \
, \llabel{060213e}
\ee
which gives the equilibrium positions for the spin of the planet, generally
known as  ``Cassini states'' \citep[e.g.][]{Henrard_Murigande_1987}.
Since $ {\cal J}_k / \nu_k \ll 1 $ (Eq.\,\ref{060213c}), the equilibrium positions
for the obliquity are then:
\be
\tan \ve \approx \Frac{{\cal J}_k}{\nu_k \pm \alpha} \ , \quad \cos \ve
\approx - \Frac{\nu_k}{\alpha} \ . \llabel{060213f}
\ee
When $ \md \alpha / \nu_k \md \ll \md \alpha / \nu \md_{crit} $, the first
expression gives states 2 and 3, while the second expression has no real roots
(states 1 and 4 do not exist).
When $ \md \alpha / \nu_k \md \gg \md \alpha / \nu \md_{crit} $, the first
expression approximates Cassini states 1 and 3, while the second one gives
states 2 and 4.
States 1, 2 and 3 are stable, while state 4 is unstable.
Although gravitational tides always decrease the obliquity (Eq.\,\ref{090520e}),
the ultimate stage of the obliquity evolution is to be captured
into a Cassini resonant state, similarly to the capture of the rotation in a
spin-orbit resonance (Eq.\,\ref{Res3}).

The complete system (Eqs.\,\ref{050820c} and \ref{060213b})
is usually not integrable  as there are several terms in expression
(\ref{050820b}), but we can look individually
to the location of each resonance. 
When the resonances are far apart, the motion will behave locally as in the
integrable case, with just the addition of supplementary small oscillations. 
However, if several resonances overlap, the motion is no longer regular and  
becomes chaotic \citep{Chirikov_1979, Laskar_1996}.
For instance, the present obliquity variations on Mars are chaotic and can vary
from zero to nearly sixty degrees \citep{Laskar_Robutel_1993,  
Touma_Wisdom_1993, Laskar_etal_2004E}.

\section{APPLICATION TO THE PLANETS}

The orbital parameters of exoplanets are reasonably well
determined from radial velocity, transit or astrometry
techniques, but exoplanets' spins remain a mystery. 

The same applies to the primordial spins of the terrestrial planets in the Solar
System, since very little constraints can be
derived from the present planetary formation models.
Indeed, a small number of large impacts at the end of the formation process will
not average, and can change the spin direction.
The angular velocities are also unpredictable, but they are usually high, $
\omega \gg n $
\citep{Dones_Tremaine_1993,Agnor_etal_1999,Kokubo_Ida_2007}, although impacts can also
form a slow rotating planet ($ \omega \sim n $) if the size of the typical accreting bodies
is much smaller than the protoplanet \citep[e.g.][]{Schlichting_Sari_2007}. 
The critical angular velocity for rotational instability is
\citet{Kokubo_Ida_2007}:
\be
\omega_{cr} \approx 3.3 \left(\frac{\rho}{3 \, \mathrm{g \, cm}^{-3}}\right)^{1/2}
\mathrm{hr}^{-1}  \ , \llabel{100220a}
\ee
which sets a maximum initial rotation periods of about $1.4$\,h, for the inner
planets of the Solar System.

For the Jovian planets in the Solar System no important mechanism susceptible of
altering the rotation rate is known, but the orientation of the axis may also
change by secular resonance with the planets \citep[e.g.][]{Correia_Laskar_2003I,Ward_Hamilton_2004,Boue_Laskar_2010}.
The fact that all Jovian planets rotate fast (Table\,\ref{Tjup}) seems to be in agreement with theoretical
predictions \citep[e.g.][]{Takata_Stevenson_1996}.

An empirical relation derived by \citet{MacDonald_1964} based
on the present rotation rates of planets from Mars to Neptune (assumed almost
unchanged) gives for the initial rotation rates 
\be
\omega_0 \propto m^{4/5} {R}^{-2} \ . \llabel{091120a}
\ee
Extrapolating for the remaining inner planets, we get initial rotation periods of about
$18.9$\,h, $13.5 $\,h, and $12.7 $\,h for Mercury, Venus and the Earth,
respectively, much faster than today's values, which in agreement with the
present formation theories.

The above considerations and expressions can also be extended to exoplanets. 
However, since many of the exoplanets are close to their host stars, 
it is believed that the spins have undergone significant tidal dissipation
and eventually reached some equilibrium positions, as it happens
for Mercury and Venus in the Solar System. 
Therefore, in this section we will first review the rotation of the terrestrial
planets, and then look at the already known exoplanets.
In particular, we will focus our attention in two classes of exoplanets, the
``Hot-Jupiters'' (fluid) and the ``Super-Earths'' (rocky with atmosphere), for
which tidal effects may play an important role in orbital and spin
evolution.

\subsection{Solar System examples}


\subsubsection{Mercury}

The present spin of Mercury is very peculiar: 
the planet rotates three times around its axis in the
same time as it completes two orbital revolutions \citep{Pettengill_Dyce_1965}.
Within a year of the discovery, the stability of this 3/2 spin-orbit resonance
became understood as the result of the solar torque on Mercury's quadrupolar
moment of inertia combined with an eccentric orbit (Eq.\,\ref{Res1}) 
\citep{Colombo_Shapiro_1966,Goldreich_Peale_1966}.
The way the planet evolved into the 3/2 configuration remained a mystery for
long time, but can be explained as the result of tidal evolution combined with
the eccentricity variations due to planetary perturbations
\citep{Correia_Laskar_2004,Correia_Laskar_2009}.
 
Mercury has no atmosphere, and the spin evolution of the planet is therefore
controlled by gravitational tidal interactions with the Sun.
Tidal effects drive the final obliquity of Mercury 
close to zero (Eq.\,\ref{090520e}), and the averaged equation for the
rotation motion near the $ p $ resonance can be written putting expressions
(\ref{Res1}) and (\ref{090515a}) together:
\be
\frac{d  \omega}{d t} = - \beta' \sin 2 \gamma
- K' \left[ \frac{\omega}{n} -  \frac{f_2(e)}{f_1(e)} \right]   \ , \llabel{eq1}
\ee
where $\gamma = \theta - p M - \phi$, $ \beta' = \beta H(p,e) $, and 
$ K' = \Kn f_1 (e) / C$.
Note that near the $p$ resonance a contribution from
core-mantle friction may also be present, but we will neglect it here 
\citep[for a full description see][]{Correia_Laskar_2009,Correia_Laskar_2010}.
The tidal equilibrium is achieved when $ d \omega / d t = 0 $,
that is, for a constant eccentricity $e$, when $ \omega / n =  f_2(e)/f_1(e) $. 
In a circular orbit ($ e = 0 $) the tidal equilibrium coincides with  
synchronization, while the equilibrium rotation rate $ \omega/n = 3/2 $
is achieved for $ e_{3/2} = 0.284927 $ (Fig.\,\ref{FigC}).

For the present value of Mercury's
eccentricity ($ e \approx 0.206 $), the capture probability in the 3/2
spin-orbit resonance is only about 7\% (Eq.\,\ref{040819b}).
However, as the eccentricity of Mercury suffers strong chaotic variations in
time due to planetary secular perturbations, 
the eccentricity can vary from nearly zero to more than 0.45,
and thus reach values higher than the critical value $ e_{3/2} = 0.284927 $.
Additional capture into resonance can then occur, at any time
during the planet's history \citep{Correia_Laskar_2004}. 

In order to check the past evolution of Mercury's spin, it is not possible to
use a single  orbital solution, as due to the chaotic behavior the motion cannot
be predicted precisely beyond a few tens of millions of years. 
A statistical study of the past evolutions of Mercury's orbit was then performed,
with the integration of 1000 orbits over 4~Gyr in the past, starting with  very
close initial conditions, within the uncertainty of the present
determinations \citep{Correia_Laskar_2004,Correia_Laskar_2009}.  
 \begin{figure}[t]
 \epsscale{0.95}
\plotone{\figpath 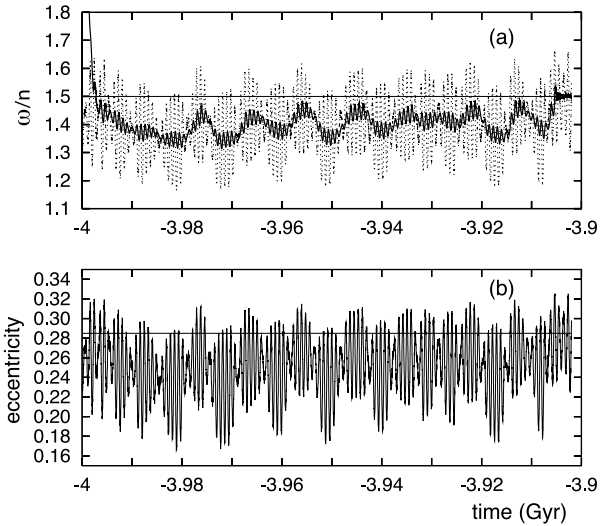}
  \caption{\small Simultaneous evolution of the rotation rate (a) and the eccentricity
  (b) of Mercury.
  In this example, there is no capture at the first encounter with the 3/2
  resonance (at $t\approx -3.9974$ Gyr). About 100 Myr later, as 
  the mean eccentricity increases, additional crossing of the 3/2 resonance 
  occurs, leading to capture with damping of the libration \citep{Correia_Laskar_2006}.
  \llabel{FNat3} }
 \end{figure}

\begin{deluxetable}{c r r r r}
\tabletypesize{\small}
\tablecaption{Possible final spin states of Venus. \llabel{4FS}}
\tablewidth{0pt}
\tablehead{
 state  & $\ve $ & $ \omega $ & $ \pt $ (days) & $ \pt_s $ (days) \\
}
\startdata
$\cF^+_0 $ & $ 0^\circ $ & $ n + \omega_s $ & $   76.83 $& $  116.75 $\\ 
$\cF^-_0 $ & $ 0^\circ $ & $ n - \omega_s $ & $- 243.02 $& $- 116.75 $\\ 
$\cF^+_\pi $ & $180^\circ$ & $ - n - \omega_s $ & $-  76.83 $& $  116.75 $\\ 
$\cF^-_\pi $ & $180^\circ$ & $ - n + \omega_s $ & $  243.02 $& $- 116.75 $\\ 
\enddata
\smallskip

There are two retrograde
states ($\cF_0^- $ and $\cF^-_\pi $) and two prograde states ($\cF_0^+ $ and
$\cF^+_\pi $). In all cases the synodic period $ \pt_s $ is the same
\citep{Correia_Laskar_2001}. 
\end{deluxetable}

For each of the 1000 orbital motion of Mercury, the rotational motion
(Eq.\,\ref{eq1}) was integrated numerically with planetary perturbations.
As $e(t)$ is not constant, $\omega(t)$ will tend towards a limit value $\tilde
\omega (t)$ that is similar to an averaged value of $(f_2/f_1)(e(t))$ and capture into
resonance can now occur more often (Fig.\,\ref{FNat3}).  
Globally, only 38.8\% of the solutions did not end into resonance,
and the final capture probability distribution was \citep{Correia_Laskar_2004}:  
$$
P_{1/1}=2.2\%,\quad  P_{3/2}=55.4\%,\quad P_{2/1}=3.6\% \ .
$$

With the consideration of the chaotic evolution of the eccentricity of Mercury,
the present $3/2$ resonant state becomes the most probable
outcome for the spin evolution.
The largest unknown remains the dissipation  factor $k_2 \Delta t$ in the
expression of $K$ (Eq.\,\ref{eq3}). 
A stronger
dissipation increases the probability of capture into the $3/2$
resonance, as $\omega/n$ would follow more closely $f_2(e)/f_1(e)$ (Fig.\,\ref{FNat3}),
while lower dissipation slightly decreases the capture probability.
The inclusion of core-mantle friction also increases the chances of capture for
all resonances \citep{Correia_Laskar_2009,Correia_Laskar_2010}.

\subsubsection{Venus} 
 
 \begin{figure}[t]
 \epsscale{0.93}
\plotone{\figpath 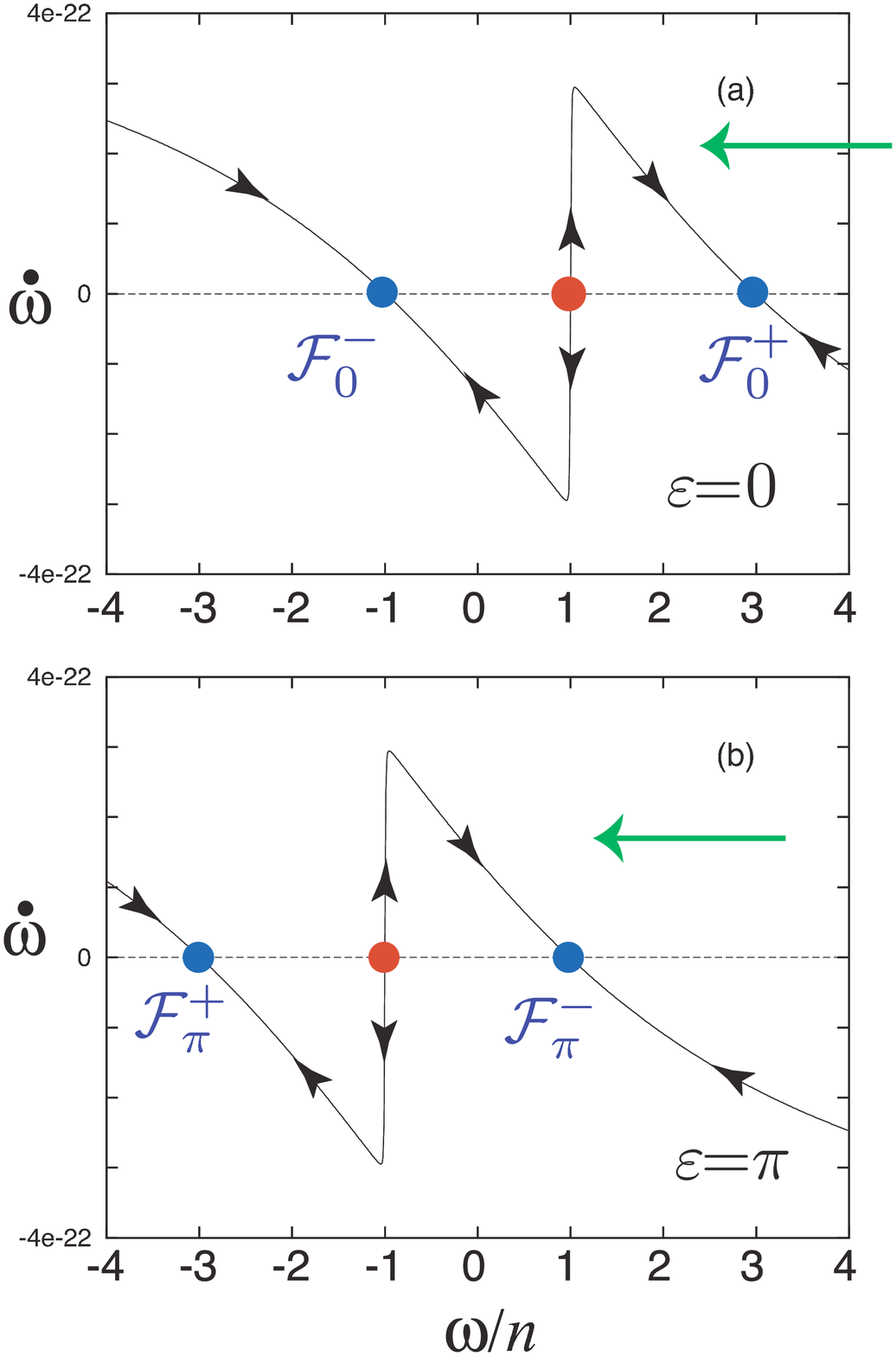}
  \caption{\small Final states for a planet with strong atmospheric thermal
  tides. The original
  equilibrium point obtained at synchronization ($ \omega / n = 1 $) 
  when considering uniquely the gravitational tides, becomes unstable, and
  bifurcates at $ \ve = 0 $ into two new stable fixed points $\cF_0^- $ and
  $\cF_0^+ $, and at $ \ve = \pi $ into $\cF_\pi^- $ and $\cF_\pi^+ $
  \citep{Correia_Laskar_2001,Correia_Laskar_2003I}. \llabel{F6}}
 \end{figure}

Venus an unique case in the Solar System: it presents a slow
retrograde rotation, with an obliquity close to 180 degrees and a 243-day
period \citep{Smith_1963,Goldstein_1964,Carpenter_1970}. 
According to planetary formation theories it is highly improbable that the
present spin of Venus is primordial, since we would expect a lower obliquity
and a fast rotating planet (Eq.\,\ref{091120a}).

The present rotation of Venus is believed to represent a steady state resulting from a
balance between gravitational tides, which drives the planet toward
synchronous rotation, and thermally driven atmospheric tides, which drives
the rotation away \cite[e.g.][]{Gold_Soter_1969}. 
The conjugated effect of tides and core-mantle friction can tilt Venus
down during the planet past evolution, but requires high values of the
initial obliquity \citep[e.g.][]{Dobrovolskis_1980,Yoder_1997}. 
However, the crossing in the past of a large chaotic zone
for the spin, resulting from secular planetary perturbations
\citep{Laskar_Robutel_1993}, can lead Venus to the present retrograde
configuration for most initial conditions
\citep{Correia_Laskar_2001,Correia_Laskar_2003I}. 

Venus has a dense atmosphere and the planet is also enough close to the Sun to
undergo significant tidal dissipation. Venus' spin evolution is then 
controlled by tidal effects (both gravitational and thermal).
Tidal effects combined can drive the obliquity either to $ \ve = 0^\circ $ or $
\ve = 180^\circ $ \citep{Correia_etal_2003}.
For the two final obliquity possibilities, the tidal components become very
simplified, with (at second order in the planetary eccentricity) a single term of 
tidal frequency $ \sigma = 2 \omega - 2 n $  for $ \ve = 0 $ and $ \sigma = 2
\omega + 2 n $ for $ \ve = \pi $ (Eq.\,\ref{V8a}). 
Combining expressions (\ref{021010c1}) and (\ref{A6a}) we can write for the
rotation rate:
\be
  \begin{array}{l l}
\left. \Frac{d \omega}{d t} \right\vert_{0} = - \Frac{3}{2}[K_g b_g({2 \omega - 2 n}) +
 K_a b_a({2 \omega - 2 n}) ] \ , \crm
\left. \Frac{d \omega}{d t} \right\vert_{\pi} = \small - \Frac{3}{2}[K_g b_g({2 \omega + 2
n}) + K_a b_a({2 \omega + 2 n}) ] \ ,
  \end{array} 
 \llabel{eqSys} 
\ee 
where $ K_g $ and $ K_a $ are given by the constant part of expressions
(\ref{021010c1}) and (\ref{A6a}), respectively. 
Let  $ f (\sigma) $ be defined as
\be 
f (\sigma) = \frac{b_a (2 \sigma)}{b_g (2 \sigma)} \ .
\ee
As $ b_\tau (\sigma) $ is an odd function of $ \sigma $ (Eq.\,\ref{V8a}),
$ f (\sigma) $ is an even function of $ \sigma $ of the form $ f (\md
\sigma \md) $.  
Thus, at equilibrium, with $ d \omega / d t = 0 $, we obtain an equilibrium
condition
\be 
f (\md \omega - \mmu n \md ) = - \frac{K_g}{K_a} \ , \llabel{eqK} 
\ee
where $\mmu=+1$ for $\ve=0$ and $ \mmu = - 1 $ for $ \ve = \pi $.
Moreover, for all commonly used dissipation models
$ f $ is monotonic and decreasing for slow rotation rates.  
There are thus only four possible values for the final rotation
rate $ \omega_f $ of Venus, given by
\be 
\md \omega_f - \mmu n \md = f^{-1} \left( -\frac{K_g}{
K_a} \right) = \omega_s \ . \llabel{eq7} 
\ee

Assuming that the present rotation of Venus corresponds to a stable
retrograde rotation, since $\omega_s > 0$ the only possibilities 
for the present rotation are $\ve=0$ and $ \omega_{obs} = n - \omega_s $, 
or $\ve = \pi $ and $\omega_{obs}=\omega_s-n$. 
In both cases, $ \omega_s = n  +  \md \omega_{obs} \md $ ($\omega_s $ is thus the
synodic frequency). 
With 
\be 
\omega_{obs}  =  {2 \pi}/{243.0185 \; \mathrm{day}}  \ ; \quad
 n  =  {2 \pi}/{224.701\; \mathrm{day} }  \ , \llabel{refPobs}
\ee
we have
\be 
\omega_s =  {2 \pi}/{116.751 \; \mathrm{day} } \ . \llabel{refPsyn} 
\ee
We can then determine all four final states for Venus (Table~\ref{4FS}). 
There are two retrograde states ($\cF^-_0 $ and $ \cF^-_\pi $) and two prograde
states ($\cF^+_0 $ and $ \cF^+_\pi $).  
The two retrograde states correspond to the observed present retrograde state of Venus 
with period $243.02$ days, while the two other states have a prograde rotation period 
of $76.83$ days.
Looking to the present rotation state of the planet, it is impossible to
distinguish between the two states with the same angular momentum (Fig.\,\ref{F6}).

In order to obtain a global view of the possible final evolutions of Venus'
spin, numerical integrations of the equations of motion for the
dissipative effects (Eqs.\,\ref{021010c1}, \ref{021010c2}, \ref{A6a} and
\ref{A6b}), with the addition of planetary perturbations 
(Eqs.\,\ref{050820c} and \ref{060213b}) were performed
\citep{Correia_Laskar_2001,Correia_Laskar_2003I}. 
In Figure\,\ref{F010531a} we  show  the possible final evolutions for a
planet starting with an initial period ranging from  3 to 12 days, with  an
increment  of 0.25 day, and initial  obliquity  from $ 0^\circ $ to $
180^\circ $, with an increment of 2.5 degrees 
(rotation periods faster than 3 days 
are excluded as they do not allow  the planet to
reach a final rotation state within the age of the Solar System).
\begin{figure}[t]
 \epsscale{1.}
\plotone{\figpath 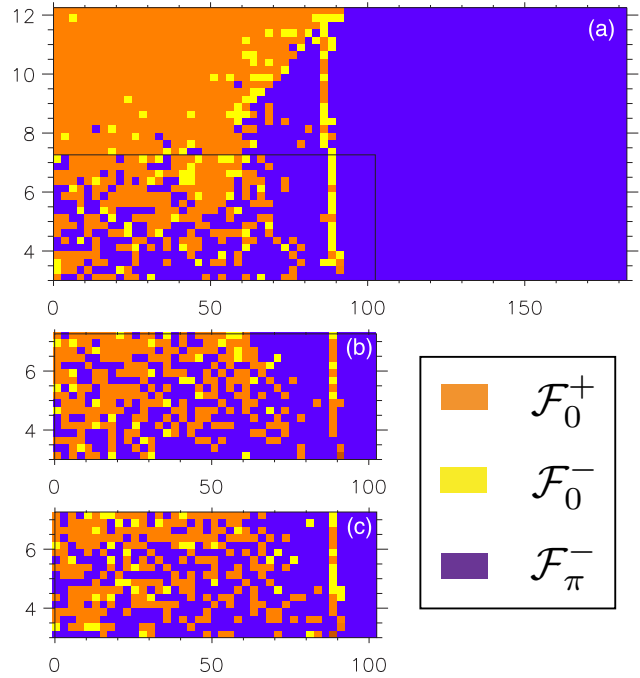}
  \caption{\small Final states of Venus' spin with planetary perturbations for initial obliquity ($\ve_i
  \in[0^\circ,180^\circ]$) and period ($ \pt_i \in [3 \, \mathrm{d},12 \, \mathrm{d}]$). 
  For high initial obliquities, the
  final evolution of Venus remains essentially unchanged since none of the
  trajectories crossed the chaotic zone. The passage through the chaotic zone is
  reflected by the scattering of the final states in the left side of the
  picture. To emphasize the chaotic behavior, in the bottom left corner of
  picture (a), additional integrations were done
  with the same initial conditions, but with a difference of $ 10^{-9} $
  in the initial eccentricity of Mars (b) and Neptune (c) \citep{Correia_Laskar_2003I}.\llabel{F010531a}}
 \end{figure}
Each color represents one of the possible final states.
For high initial obliquities, the spin of Venus always evolves into the
retrograde final state $ \cF^-_\pi $.
It is essentially the same evolution as without planetary perturbations, 
since none of the trajectories encounters a chaotic zone for the obliquity
\citep{Laskar_Robutel_1993}. 
However, for evolutionary paths starting with low initial obliquities, we can
distinguish two different zones: one zone corresponding to slow 
initial rotation periods ($ \pt_i > 8 $~day) where the prograde rotation final state $
{\cal F}_0^- $ is prevailing, and another zone for faster initial rotation
periods ($ \pt_i < 8 $~day), where we find a mixture of the three attainable final states, $ \cF^+_0
$, $ \cF^-_0 $ and $ \cF^-_\pi $. 
To emphasize the chaotic behavior, we integrated  twice more the zone with $
\pt_i < 8 $~day, 
with a difference of $ 10^{-9} $ in the initial eccentricity of Mars
(Fig.\,\ref{F010531a}b), and   with a difference of $ 10^{-9} $ in the initial 
eccentricity of Neptune (Fig.\,\ref{F010531a}c).
The passage through the chaotic zone is reflected by the scattering of the final states
in the left hand side of the picture.

\subsubsection{The Earth and Mars}

Contrary to Mercury and Venus, the Earth and Mars are not tidally evolved. 
For Mars, the tidal dissipation from the Sun is negligible.
For the Earth, tidal dissipation is noticeable due to the presence of
the Moon, but the Earth's spin is still far from the equilibrium 
\citep[e.g.][]{Surgy_Laskar_1997}. 
Nevertheless, the spin axis of both planets is subjected to planetary
perturbations and thus present some significant variations (Sect.~2.5).

In the case of Mars, the presence of   numerous secular resonances 
of the kind $\dot \psi = -\nu_k$ (Eq.\,\ref{050820c}) induce large chaotic
variations  in the obliquity, which can evolve between 0 and $60$ degrees
(Fig.\,\ref{Figmars}).
At present, the obliquity of Mars is very similar to the obliquity of the Earth,
which is a mere coincidence. 
Indeed, the obliquity of Mars has most certainly reached values larger than 45
degrees in the past \citep{Laskar_etal_2004M}.
The high obliquity periods  led to large climatic changes on Mars, with possible
occurrence of large scale ice cycles where the polar caps are sublimated during
high obliquity stages and the ice is  deposited in the equatorial regions 
\citep{Laskar_etal_2002,Levrard_etal_2004,Levrard_etal_2007JGR}.

\begin{figure}[t]
 \epsscale{0.95}
\plotone{\figpath 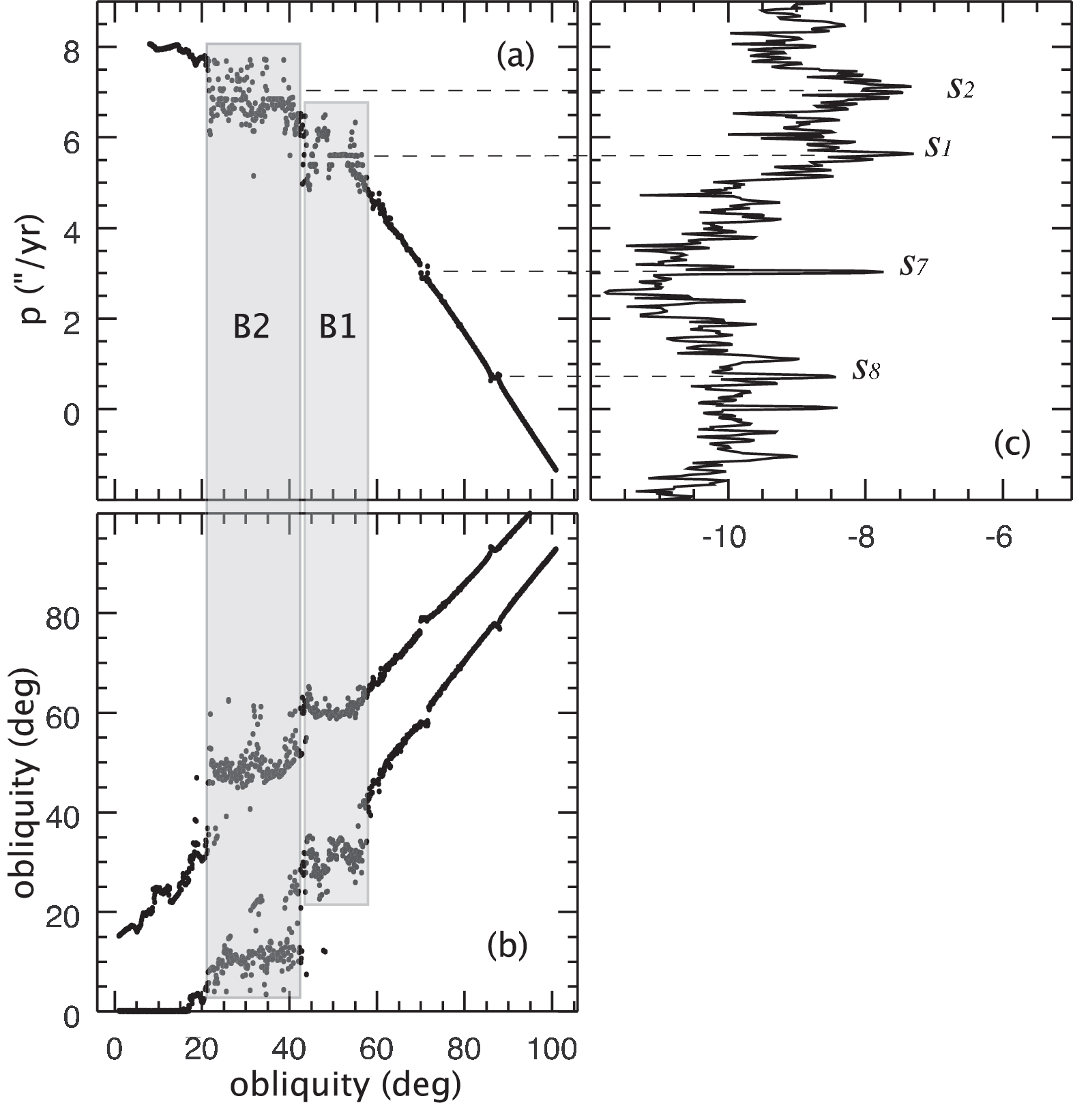}
  \caption{\small Frequency analysis of Mars' obliquity.
  (a)  The frequency map is obtained by reporting in the ordinate the
 precession frequency value obtained for 1000 integrations  
 over 56 Myr for the different values of the initial obliquity (abscissa).
 A large chaotic zone is visible, ranging from $0^\circ$ to
about $60^\circ$, with two distinct chaotic zones, $B1$ and $B2$.
(b) Maximum  and minimum values of  the obliquity reached over
56 Myr. 
In (c), the power spectrum of the orbital forcing term 
(Eq.\,\ref{050820b}) is given in logarithmic scale, showing the correspondence
of the chaotic zone with the main secular frequencies $s_1,s_2,s_7,s_8$.
\citep{Laskar_etal_2004M}.  \llabel{Figmars}}
\end{figure}

\begin{figure}[t]
 \epsscale{1.}
\plotone{\figpath 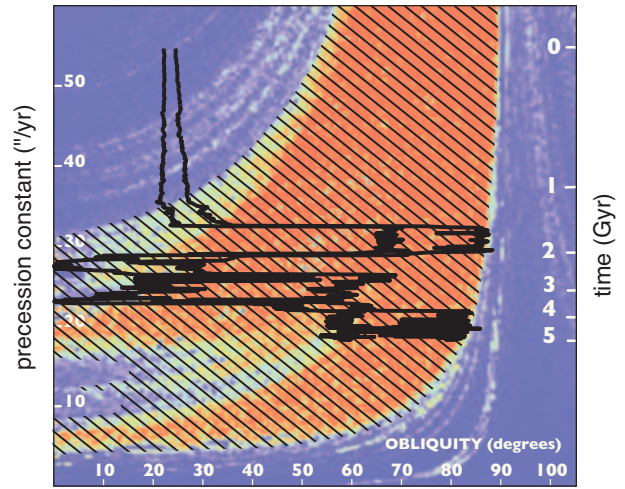}
\caption{\small Example of possible evolution of the Earth's obliquity for 5~Gyr
in the future, due to tidal dissipation in the Earth-Moon system.
The background of the figure 
is obtained  as the result of a stability analysis 
on about 250\,000 numerical integrations 
of the obliquity of the Earth under planetary perturbations for 36~Myr, for
various values of the initial obliquity of the Earth ($x$-axis) and 
various values of the precession constant (left $y$-axis).
We observe a very large chaotic zone (with stripes on the figure)
resulting from overlap of orbital secular resonances
\citep{Laskar_etal_1993N,Laskar_Robutel_1993}.
The numerical integration is then conducted over 5~Gyr years in the future 
for the obliquity of the Earth, including tidal dissipative effects.
The two bold curves correspond to the minimum and maximum values  reached by the obliquity
and the time scale is given in the right $y$-axis.
As long as the orbits stays in the regular region,
 the motion suffers only small 
(and regular) variations. As soon as the orbit enters the chaotic
zone, very strong variations of the obliquity are observed, which wanders in all the 
chaotic zone, and very high values,  close to $90$ degrees, are reached
\citep{Surgy_Laskar_1997}.   \llabel{Figearth} } 
 \end{figure}

In the case of the Earth, the precession frequency is not in resonance with any
orbital secular frequency ($\dot \psi \neq \nu_k $).  
The obliquity of the Earth is then only subject to  small 
oscillations of about 1.3 degrees around the mean value (23.3 degrees) 
with main periodicities around 40\,000 years \citep{Laskar_etal_2004E}.
The small obliquity variations are nevertheless sufficiently important to 
induce substantial changes in the insolation received in summer 
in high latitude regions 
on the Earth, and they are imprinted in the geological stratigraphic 
sequences \citep{Hays_etal_1976,Imbrie_1982}.

Due to tidal dissipation in the Earth-Moon system, the Moon is 
moving away from Earth at a 3.8~cm/yr rate \citep{Dickey_etal_1994}, and the
rotation rate of the Earth is slowing down (Eqs.\,\ref{090515b} and \ref{090515a},
respectively).
As a consequence, the torque exerted on the equatorial bulge of the Earth decreases 
and thus also the Earth precession frequency (Eqs.\,\ref{050109a} and
\ref{061120a}).
Using the present dissipation parameters of the Earth, 
\citet{Surgy_Laskar_1997} found that after 1.5~Gyr, the spin of the Earth will 
enter a large chaotic zone of overlapping orbital secular resonances. 
From then, the Earth spin axis will evolve in a wildly 
chaotic way, with a  possible range from 0 to nearly 90 degrees (Fig.\,\ref{Figearth}).

The main difference between the Earth and Mars is thus due to the presence of
the Moon, whose gravitational torque on the equatorial bulge of the 
Earth prevents the spin axis to evolve in a largely chaotic state. 
In absence of the Moon, the behavior of the spin axis of the Earth and Mars
would be identical \citep{Laskar_etal_1993N,Laskar_Robutel_1993}. 

Depending on the orbital configuration of the exoplanetary systems, we thus expect 
to find planets that would be  either in a chaotic state as Mars or the moonless Earth, 
or in a regular state as the Earth with the Moon. 
It should be stressed, however, the presence of a large satellite is not
mandatory in order to stabilize the spin axis. 
Since the stability of the axis is very important for the exoplanet climate,
planetary perturbations should be taken into consideration when searching for 
another Earth-like environments.

\begin{deluxetable}{c c c c c}
\tabletypesize{\small}
\tablecaption{Constants for the Solar System outer planets.
\llabel{Tjup}}
\tablewidth{0pt}
\tablehead{
 quantity  & Jupiter & Saturn & Uranus & Neptune \\
}
\startdata
$ P_0 $ (h)         & 9.92 & 10.66 & 17.24 & 16.11 \\ 
$ \rho$ (g/cm$^3$)  & 1.33 &  0.69 &  1.32 &  1.64 \\ 
$C/MR^2 $     & 0.25 &  0.21 &  0.23 &  0.24 \\ 
$k_2$               & 0.49 &  0.32 &  0.36 &  0.41 \\ 
$Q \ (\times 10^4)$ &$\sim$3&$\sim$2&1$\sim$3& 1$\sim$30 \\ 
\enddata
\smallskip

\citep{Yoder_1995,Veeder_etal_1994, Dermott_etal_1988, Tittemore_Wisdom_1990,
Banfield_Murray_1992}.
\end{deluxetable}

\subsection{``Hot-Jupiters''}

One of the most surprising findings concerning exoplanets was the discovery of
several giant planets with periods down to 3~days,
that were designated by ``Hot-Jupiters'' \citep[e.g.][]{Santos_etal_2005}.
Many of the ``Hot-Jupiters'' were simultaneously
detected by transit method and radial Doppler shift, which allows the
direct and accurate determination of both mass and radius of the exoplanet. 
Therefore, ``Hot-Jupiters'' are amongst the better characterized planets
outside the Solar System.

Due to the proximity of the host star, ``Hot-Jupiters'' are almost
certainly tidally evolved.
Given the large mass of ``Hot-Jupiters'', they may essentially be
composed of an extensive Hydrogen atmosphere, similar to Jupiter or Saturn.
As a consequence, despite the presence of an inner metallic core,
``Hot-Jupiters'' can be treated as fluid planets, and we may adopt a viscous
model for the tidal dissipation (Eqs.\,\ref{090515a}, \ref{090520d}).
Thermal atmospheric tides may also be present
\citep[e.g.][]{Arras_Socrates_2010b,Arras_Socrates_2010a,Goodman_2010}, but we
did not take thermal tides into account, as gravitational tides are so strong
for ``Hot-Jupiters'', that they probably rule over all the remaining 
effects (see Sect.~3.3).

\subsubsection{Rotational synchronization}

The effect of gravitational tides over the obliquity is to straighten the spin axis
(Eq.\,\ref{090520e}), so we will adopt $ \ve = 0^\circ $ for the obliquity.
Assuming that the eccentricity and the semi-major axis of the planet are constant, we
can derive from expression (\ref{090515a}):
\be
\frac{\omega}{n} = \frac{f_2(e)}{f_1(e)} + \left( \frac{\omega_0}{n} -
\frac{f_2(e)}{f_1(e)} \right) \exp (- t / \tau_{eq}) \ , \llabel{091123a}
\ee
where $ \tau_{eq}^{-1} = K f_1(e) / C $ is the characteristic 
time-scale for fully despinning the planet.

As for Mercury, the final equilibrium rotation driven by tides ($ t \rightarrow
+ \infty $) is given by the
equilibrium position $ \omega_e / n = f_2(e)/f_1(e) $, which is different from synchronous
rotation if the eccentricity is not zero (Fig.\,\ref{FigC}).
Unlike Mercury, because ``Hot-Jupiters'' are assumed to be fluid, they should
not present much irregularities in the internal structures. 
Therefore $ (B-A)/C  \approx 0 $ and we do not expect ``Hot-Jupiters'' to be
captured in spin-orbit resonances.
Indeed, determination of second-degree harmonics of
Jupiter and Saturn's gravity field from Pioneer and Voyager tracking data
\citep{Campbell_Anderson_1989} provided a crude estimate of the
$(B-A)/C$ value lower than $\sim 10^{-5}$ for Saturn and $\sim 10^{-7}$ for
Jupiter.
The $(B-A)/C$ values for Jupiter and Saturn are more than one order of magnitude
smaller than the Moon's or Mercury's value, leading to insignificant chances of
capture.  
Furthermore, the detection of an equatorial asymmetry is questionable.
If an equatorial bulge originates from local mass inhomogeneities driven by
convection, it is probably not permanent and must have a more
negligible effect if averaged spatially and temporally. 

The time required for dampening the rotation of the planet depends on
the dissipation factor $ k_2 \Delta t $ (Eq.\,\ref{eq3}). 
Assuming that ``Hot-Jupiters'' are similar to the Solar System giant planets, we
can adopt $k_2 = 0.4 $,  
and a range for $Q $ from $10^4$ to $10^5$ (Table\,\ref{Tjup}).
The $Q$-factor and the time lag $ \Delta t $ can be relied using expressions
(\ref{090419d}) and (\ref{090419e}):
\be
Q^{-1} \approx \sigma \Delta t \ . \llabel{100327a}
\ee
Since we are using a viscous model, for which $ \Delta t $ is made
constant, $Q$ will be modified across the evolution as $Q$ is inversely
proportional to the tidal frequency $\sigma$.
The $Q$-factor for the Solar System gaseous planets is measured for their present
rotation states, which correspond to less than one day (Table\,\ref{Tjup}).
We may then assume that exoplanets should present identical $Q$
values when they were rotating as rapidly as Jupiter, that is, $ Q_0^{-1} =
\omega_0 \Delta t $.
For $ Q_0 = 10^4 $ and $ \omega_0 = 2 \pi / 10 $\,h, we compute a constant $
\Delta t \approx 0.57 $\,s.

We have plotted in Figure~\ref{LC3}, all known
exoplanets, taken from The Extrasolar Planets Encyclopedia ({\sf
http://exoplanet.eu/}), that could have been tidally evolved. 
We consider that exoplanets are fully evolved if
their rotation rate, starting with an initial period of 10~h, is
dampened to a value such that $ \md \omega / n - f_2(e)/f_1(e) \md < 0.01$. 
The curves represent the planets that are tidally evolved in a given time
interval ranging from 0.001~Gyr to 10~Gyr
. 
Figure~\ref{LC3} allows us to check whether the planet should be fully evolved.  
For a Solar type star, we can
expect that all exoplanets that are above the 1~Gyr curve have already reached the
equilibrium rotation $ \omega_e $.
On the other hand, exoplanets that are below the 10~Gyr curve are probably not
yet fully tidally evolved.
As expected, all planets in circular orbits with $ a < 0.05 $\,AU are tidally
evolved. 
However, we are more interested on exoplanets further from the star with non
zero eccentricity which are tidally evolved, since the rotation period is not
synchronous, but given by expression (\ref{090520a}).  
For instance, for the planet around HD\,80606, the orbital period is 111.7~days,
but since $ e = 0.92 $ we predict a rotation period of about 1.9~days. 

\begin{figure}[t]
 \epsscale{1.}
\plotone{\figpath 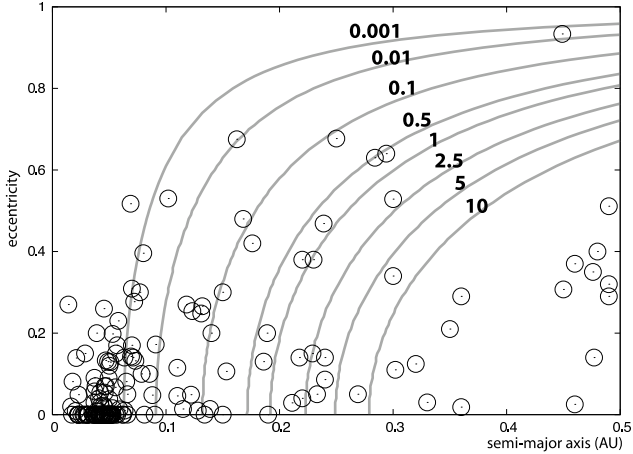}
  \caption{\small Tidally evolved exospins with $Q_0 = 10^4$ and initial rotation period $
  P_0 = 10$\,h. The labeled curves denote (in Gyr) the time needed by the
  rotation to reach the equilibrium (time-scales are linearly proportional to
  $Q_0$). We assumed Jupiter's geophysical parameters  for all planets
  (Table\,\ref{Tjup}) \citep[updated from][]{Laskar_Correia_2004}. \llabel{LC3}
  }  
 \end{figure}

\subsubsection{Cassini states}

Until now we have assumed that the final obliquity of the planet is zero
degrees.
However, \citet{Winn_Holman_2005} suggested that high obliquity values could be
maintained if the planet has been trapped in a Cassini state resonance
(Eq.\,\ref{060213e}) since the early despinning process.
For small amplitude variations of the eccentricity and inclination, the
equilibrium positions for the obliquity are given by expression (\ref{060213f}).
Unless $\alpha=|\nu_k|$, the state 1 is close to $0^{\circ}$, and state 3
is close to $180^\circ$.
We thus focus only on state~2 that may maintain a significant obliquity.

To test the possibility of capture in the high oblique Cassini state 2, we can
consider a simple scenario where a ``Hot-Jupiter'' forms at a large orbital distance
($\sim$ several AUs) and migrates inward to the current position ($\sim 0.05$~AU).
Before the planet reaches typically $\sim 0.5$~AU, tidal effects do not affect
the spin evolution, but the reduction in the semi-major axis increases the
precession constant (Eq.\,\ref{061120a}), so that the precession frequency $ \dot
\psi $ may become resonant with some orbital frequencies $\nu_k$
(Eq.\,\ref{060213b}). 
The passage through resonance generally causes the obliquity to change
\citep{Ward_1975,Ward_Hamilton_2004,Hamilton_Ward_2004,Boue_etal_2009}, rising
the possibility that the obliquity has a 
somewhat arbitrary value when the semi-major axis attains $\sim 0.5$~AU.
\begin{figure*}[t]
 \epsscale{2}
\plotone{\figpath 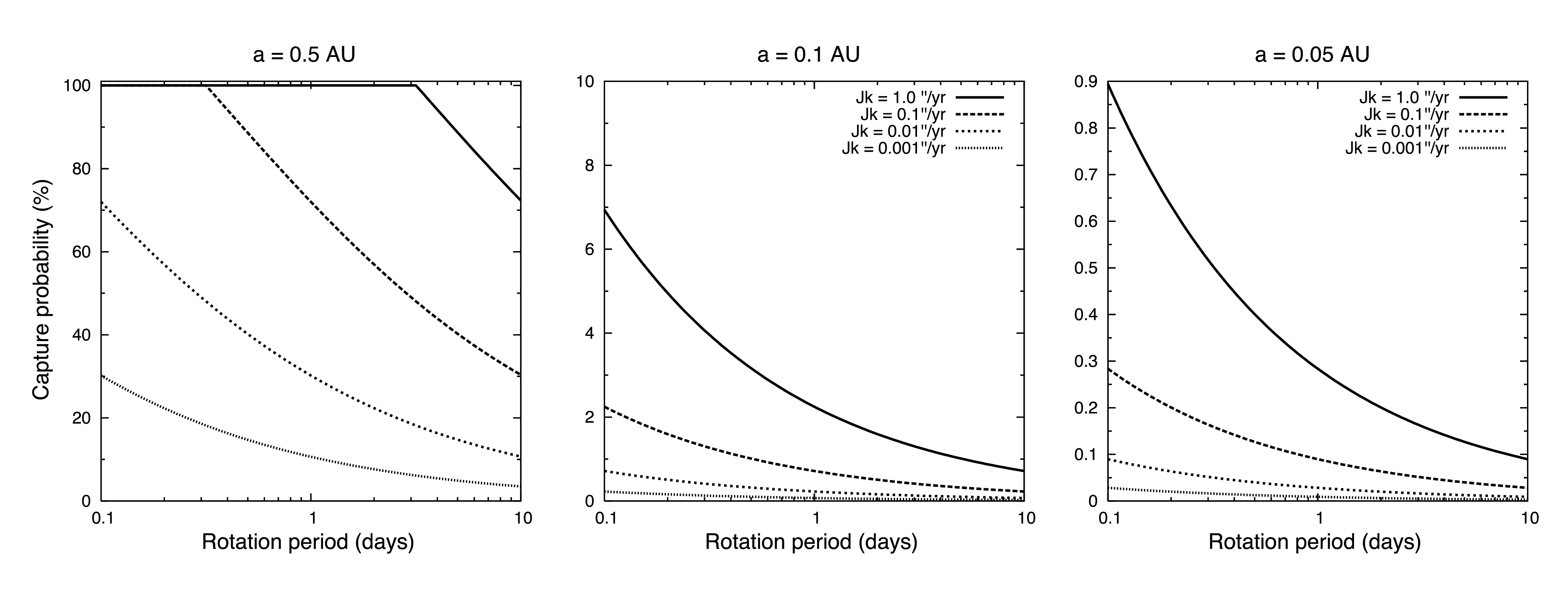}
  \caption{\small Obliquity capture probabilities in resonance
  under the effect of gravitational tides, as a function of the rotation period
  at a) 0.5~AU b) 0.1~AU c) 0.05~AU. ${\cal J}_k$ is the amplitude of the
secular orbital perturbations and $\nu_k = -10.0$''/yr 
\citep{Levrard_etal_2007}.
\llabel{Fig2}}
 \end{figure*}
 
Tidal effects become efficient for $a <  0.5$~AU and drive the obliquity to an
equilibrium value $ \cos \ve \approx 2 n (1 + 6 e^2) / \omega $ (Eq.\,\ref{090520d}).
For initial fast rotation rates ($ \omega \gg n $), the equilibrium obliquity
tends to $90^\circ $. 
As the rotation rate is decreased by tides, the equilibrium obliquity is reduced
to zero degrees (Eq.\,\ref{090520e}). 
It is then possible that the obliquity crosses several resonances
(one for each frequency $\nu_k$) in both ways (increasing and decreasing
obliquity), and that a capture occurs.
Inside the resonance island, the restoring
torque causes the obliquity to librate with amplitude
\citep{Correia_Laskar_2003I}:  
\be
\cos \ve_2 \pm \Delta \cos \ve_2 \approx - \frac{\nu_k}{\alpha} \pm 2
\sqrt{\frac{{\cal J}_k}{\alpha} \sqrt{1-\frac{\nu_k^2}{\alpha^2}}} \ .
\llabel{050820d} 
\ee
Using a linear approximation of the tidal torque (Eq.\,\ref{090520d})
around the resonant obliquity $\ve_2$,
the probability of capture in the Cassini state 2 can be estimated from the
analytical approach for spin-orbit resonances (Eq.\,\ref{Res3}), with
\citep{Levrard_etal_2007}
\be
\frac{\Delta \omega}{\pi V n}
= \left[\frac{(1-3 \cos^2\varepsilon_2)\frac{\omega}{n}+2 \cos\varepsilon_2}
{\pi\,\sin^2 \varepsilon_2\,(2-\cos\varepsilon_2\,\omega/n)}\right] \Delta \cos
\ve_2 \ .
\ee
In Figure\,\ref{Fig2}, we plotted
the capture probabilities at 0.05, 0.1 and 0.5 AU as a function of the rotation
period for different amplitudes (${\cal J}_k$) and frequencies ($\nu_k$) characteristic
of the Solar System \citep{Laskar_Robutel_1993}.
As a reasonable example, we choose $\nu_k= -10$''/yr, 
but the results are not affected by changes on this value.
Assuming an initial rotation period of 12~h,
the capture is possible and even unavoidable if ${\cal J}_k > $ 0.1''/yr at 0.5 AU. 
On the contrary, the chances of capture at 0.05~AU are negligible ($<$ 1\%)
because a decrease in the semi-major axis leads to an increase in the precession
constant and reduces the width of the resonance (Eq.\,\ref{050820d}).
Theoretical estimations can be compared with numerical simulations
\citep{Levrard_etal_2007}.
To that purpose, the spin equations (Eqs.\,\ref{050820c}, \ref{060213b}) were
integrated in the presence of tidal effects (Eqs.\,\ref{090515a}, \ref{090520d})
considering 1000 initial precession angles equally distributed over $0-2\pi$ for
each initial obliquity. 
Statistics of capture were found to be in good agreement with previous theoretical estimates.

To test the influence of migration on the capture stability, additional numerical
simulations were performed for various initial obliquities and secular
perturbations over typically $\sim 5 \times 10^7$~yr. 
The migration process was simulated by an exponential decreasing
of the semi-major axis towards 0.05 AU with a $10^5-10^7$~yr time scale.
The obliquity librations were found to be significantly shorter than spin-down
and migration time scales so that the spin trajectory follows an ``adiabatic
invariant'' in the phase space. 
Nevertheless, expression (\ref{090520d}) indicates that the tidal torque
dramatically increases both with spin-down and inward migration processes ($  d
\ve / dt \propto a^{-15/2} \omega^{-1} $). 
If the tidal torque exceeds the maximum possible restoring torque
(Eq.\,\ref{050820c}), the resonant
equilibrium is destroyed (the evolution is no longer adiabatic).
For a given semi-major axis, the stability condition requires then
that the rotation rate must always be larger than a threshold value $\omega_{crit.}$,
which is always verified if $\omega_{crit.} < f_2(e)/f_1(e)$.
The stability condition can be simply written as \citep{Levrard_etal_2007}
\begin{equation}
\tan(\ve) < {\cal J}_k \times \tau_{eq} \ ,
\llabel{cond}
\end{equation}
where $\tau_{eq}$ is the time scale of tidal despinning (Eq.\,\ref{091123a}).
It then follows that the final obliquity of the planet cannot be too large,
otherwise the planet would quit the resonance.
For instance, taking ${\cal J}_k=1$~''/yr at 0.05~AU (the highest value in
Fig.\,\ref{Fig2}), we need an obliquity $\ve < 21^{\circ}$. 
For the more realistic amplitude ${\cal J}_k=0.1$~''/yr, the resonant
obliquity drops to $\ve < 2^{\circ}$.
Such a low resonant obliquity at 0.05~AU is highly unlikely, because,
according to expression (\ref{060213f}), the resonant state requires
very high values of the orbital secular frequencies ($|\nu_k| > 7.2
\times 10^6$ ''/yr). 
At 0.5 and 0.1~AU, critical obliquity values are respectively 83$^\circ$ and
41$^{\circ}$ and require more reasonable orbital secular frequencies so that a
stable capture is possible.   
In numerical simulation, the stability criteria for the final obliquity
(Eq.\,\ref{cond}) is empirically retrieved with an excellent agreement
\citep{Levrard_etal_2007}.
When the obliquity leaves the resonance, the obliquity ultimately
rapidly switches to the resonant stable Cassini state 1, which tends to
$0^{\circ}$ (Eq.\,\ref{060213f}).
We then conclude that locking a ``Hot-Jupiter'' in an oblique Cassini state
seems to be a very unlikely scenario.

\subsubsection{Energy balance}

Tidal energy is dissipated in the planet at the expense of
the rotational and orbital energy so that
$\dot E = - C \omega \dot \omega - \dot a (G m_\star m)/(2 a^2) $.
Replacing the equilibrium rotation given by equation (\ref{090520a}) in
expression (\ref{090515b}) we obtain for the tidal energy:
\begin{equation}
\dot{E} =
K n^2 \left[f_3(e)-\frac{f_2^2(e)}{f_1(e)} \frac{2 x^2}{1+x^2} \right] \ ,
\label{tidal_energy} 
\end{equation}
or, at second order in eccentricity,
\begin{equation}
\dot{E} = \frac{K n^2}{1+\cos^2 \ve}
\left[\sin^2 \ve +e^2\left(7+16\sin^2 \ve \right) \right] \ ,
\label{En_DL}  
\end{equation}
which is always larger than in the synchronous case 
\citep[e.g.][]{Wisdom_2004}. 
In Figure\,\ref{Fig1Benj} the rate of tidal heating within
a non-synchronous and synchronous planet as a function of
the eccentricity ($0< e < 0.25$) is compared for two different obliquities
\citep{Levrard_etal_2007}.
The ratio between the tidal heating in the two situations is an increasing
function of both eccentricity and obliquity. 
For $e \approx 0$, as observed for ``Hot-Jupiters'', the ratio may reach $\sim$ 1.3
and 2.0 at respectively $45^{\circ}$ and $90^{\circ}$ obliquity, not being
significantly modified at larger eccentricity.

\begin{figure}[t]
 \epsscale{1.}
\plotone{\figpath 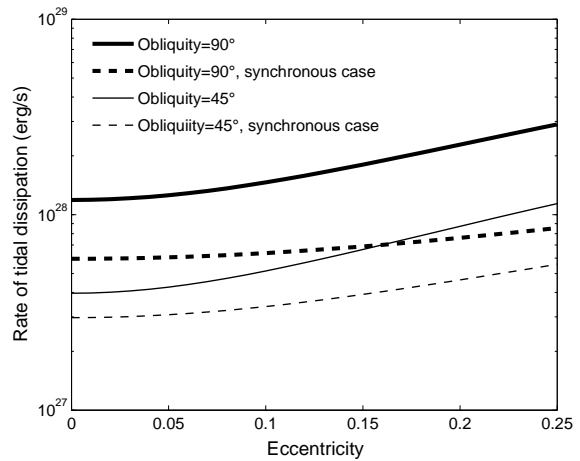}
  \caption{\small Rate of tidal dissipation within HD\,209458\,b as
   a function of the eccentricity for $45^{\circ}$ (solid thin line) and
   $90^{\circ}$ (solid thick line) obliquity. The synchronous case
   is plotted with dashed lines for comparison. The dissipation factor
   $Q_0 / k_2$ is set to $10^6$ \citep{Levrard_etal_2007}.
   \llabel{Fig1Benj}}  
 \end{figure}

We then conclude that planets in eccentric orbits and/or with high obliquity
dissipate more energy than planets in synchronous circular orbits.
This may explain why some planets appear to be more inflated than initially
expected \citep[e.g.][]{Knutson_etal_2007}.
A correct tidal energy balance must then take into account the present spin and
eccentricity of the orbit.

\subsubsection{Orbital circularization}

In Sect.~2.2.4 we saw that under tidal friction the spin of the planet attains
an equilibrium position faster than the orbit.
As a consequence, we can use the expression of the equilibrium rotation rate
(Eq.\,\ref{090520a}) in the semi-major axis and eccentricity variations and find
simplified expressions (Eqs.\,\ref{090522a} and\,\ref{090522b}).
Combining the two equations we get
\be
\frac{d a}{a} = \frac{2 e \, de}{(1-e^2)}  \ , \llabel{100210a}
\ee
whose solution is given by 
\be
a = a_f (1-e^2)^{-1} \ . \llabel{100210z}
\ee 
Replacing the above relation in expression (\ref{090522b}) we find a
differential equation that rules the eccentricity evolution:
\be
\dot e = - K_0 f_6(e) (1-e^2)^9 e \ , \llabel{100210b} 
\ee
where $ K_0 $ is a constant parameter:
\be
K_0 = \Delta t \frac{21 k_2 G m_\star^2 R^5}{2 \mu a_f^8} \ .
\llabel{100210c} 
\ee
The solution of the above equation is given by
\be
F(e) = F(e_0) \ei^{- K_0 t} \ , \llabel{100210d} 
\ee
where $ F(e) $ is an implicit function of $ e $, which converges to zero as $ t
\rightarrow + \infty $.
For small eccentricities, we can neglect terms in $ e^4 $ and $ F(e) = e
|7-9 e^2|^{-1/2} $.
The characteristic time-scale for fully dampening the eccentricity of the orbit
is then $ \tau_{orb} \sim 1 / K_0 $, and
the ratio between the spin and orbital time-scales
\be
\frac{\tau_{eq}}{\tau_{orb}} \sim \left( \frac{R}{a_f} \right)^2 \ .
\llabel{100210e} 
\ee
Since $ a_f = a (1-e^2) $, for initial very eccentric orbits the two
time-scales become comparable.

We have plotted in Figure~\ref{LC4}, all known
exoplanets, taken from The Extrasolar Planets
Encyclopedia ({\sf http://exoplanet.eu/}), whose orbits could have been tidally
evolved.
We consider that they are fully evolved if the eccentricity is
dampened to a value $ e < 0.01$. 
The curves represent the time needed to damp the eccentricity starting with the
present orbital parameters, for time
intervals ranging from 0.001~Gyr to 100~Gyr.
Figure~\ref{LC4} allows us to simultaneously check whether the planet is tidally
evolved, and the time needed to fully damp the present eccentricity.
All planets in eccentric orbits experience stronger tidal effects
because the planet is close to the star at the periapse.
As a consequence, tidal friction can, under certain conditions, be an important
mechanism for the formation of ``Hot-Jupiters'' (see Sect.~3.2.5).

\begin{figure}[t]
 \epsscale{1.}
\plotone{\figpath 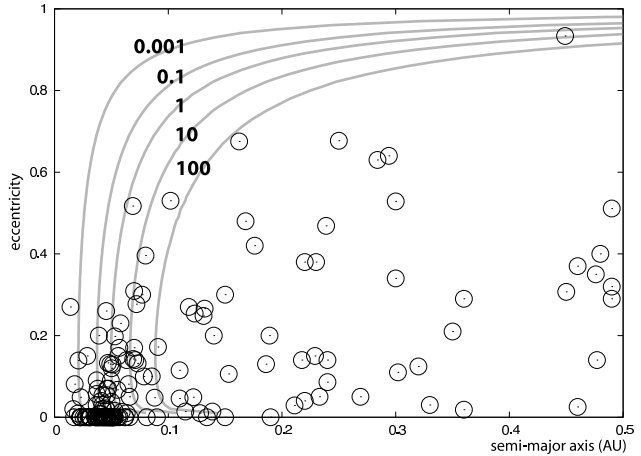}
  \caption{\small Tidally evolved orbits of exoplanets with $Q_0 = 10^4$. The
  labeled curves denote 
  (in Gyr) the time needed to circularize the orbits of the planets ($e < 0.01
  $) (time-scales are linearly proportional to $Q_0$). We assumed Jupiter's
  geophysical parameters for all planets (Table\,\ref{Tjup}).   \llabel{LC4}} 
 \end{figure}

According to Figure~\ref{LC4}, a significant fraction of exoplanets that are
close to the host star ($ a < 0.1 $~AU) still present eccentricities up to 0.4,
although tidal effects should have already damped the eccentricity to zero.
Observational errors and/or weaker tidal dissipation ($ Q_0 \gg 10^4 $) can be
a possible explanation, but they will hardly justify all the observed situations.
A more plausible explanation is that the eccentricity of the exoplanet is being
excited by gravitational perturbations from an outer planetary companion (Sect.~2.5).
Indeed, the eccentricity of a inner short-period planet can be
excited as long as its (non-resonant) outer companion's eccentricity is non-zero.
\citet{Mardling_2007} has shown that the eccentricity of the outer planet will
decay on a time-scale which depends on the structure of the inner planet, and
that the eccentricities of both planets are damped at the same rate, controlled
by the outer planet (Fig.\,\ref{Mard07}).
The mechanism is so efficient that the outer planet may be an Earth-mass
planet in the ``{\it habitable zones}'' of some stars.  
As a consequence, the evolution time-scale for both eccentricities can attain
the Gyr instead of Myr, which could explain the current observations of non-zero
eccentricity for some ``Hot-Jupiters''.

\begin{figure*}[t]
 \epsscale{2.}
\plotone{\figpath 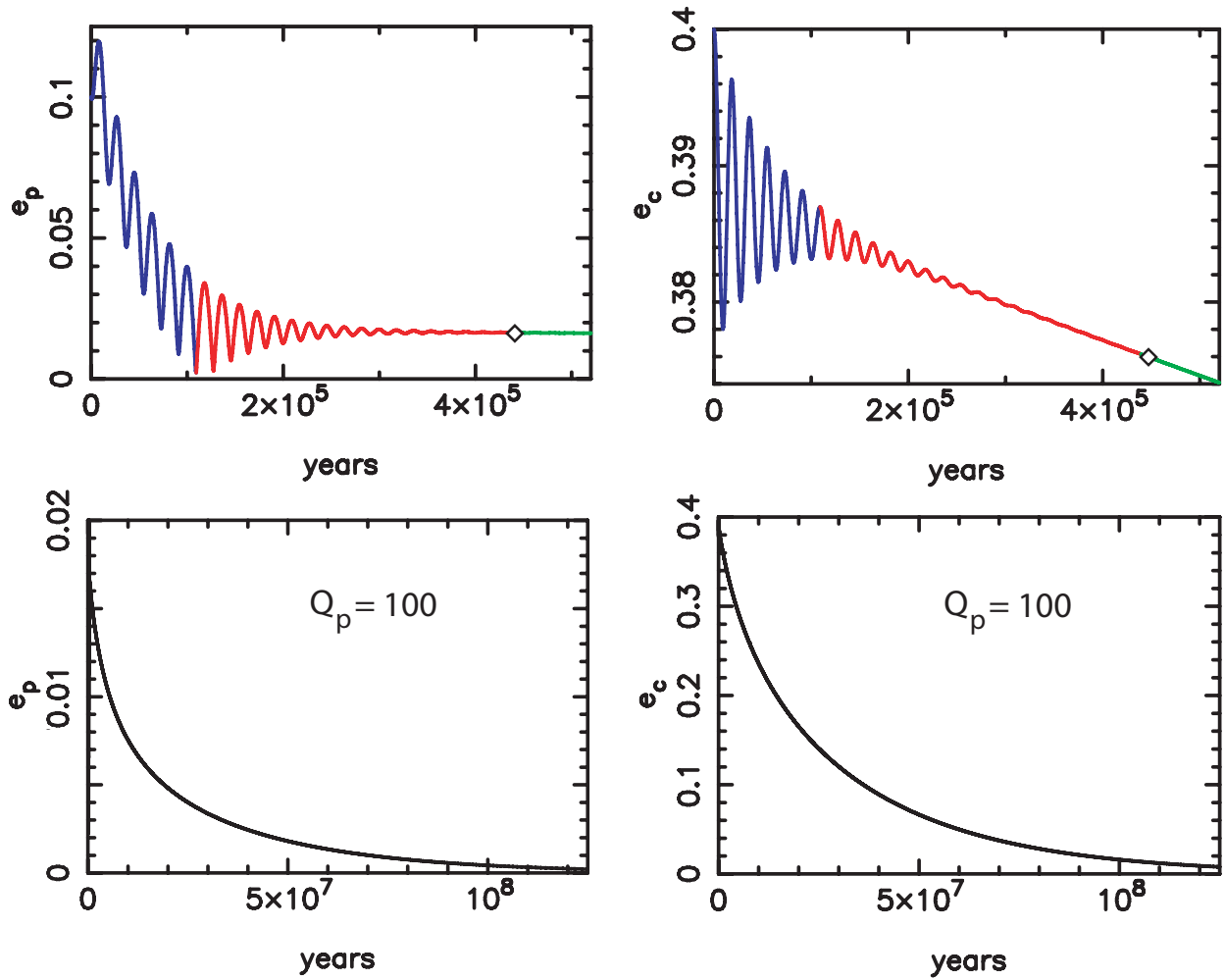}
  \caption{\small Tidal evolution of the eccentricities of planet HD\,209458\,b
  ($e_p$) perturbed by a $0.1 M_{\rm Jup} $ companion at 0.4~AU ($e_c$). The dissipation
  factor $Q_p = 100$ of the observed planet is set artificially low in order to
  illustrate the damping process (time-scales are linearly proportional to
  $Q_p$). Top figures show the first stages of the evolution. The change in
  grayscale shows the transition of the eccentricity from the circulation
  phase to the libration phase. The diamond represents the moment where the
  eccentricity librations are damped. Bottom figures show the final evolution of
  the eccentricities. The presence of a companion result that $e_p$ decays at the
  same rate of $e_c$, while the dissipation rate for $e_c$ is controlled by $Q_p$ and
  not by $Q_c$ \citep{Mardling_2007}. \llabel{Mard07}}
 \end{figure*}

\subsubsection{Kozai migration}

In current theories of planetary formation, the region within 0.1~AU of
a protostar is too hot and rarefied for a Jupiter-mass planet to form, so
``Hot-Jupiters'' likely form further away and then migrate inward.
A significant fraction of ``Hot-Jupiters'' has been found in systems of binary
stars \citep[e.g.][]{Eggenberger_etal_2004}, suggesting that the stellar companion may play an
important role in the shrinkage of the planetary orbits.
In addition, close binary star systems (separation comparable to the stellar
radius) are also often accompanied by a third star.
For instance, \citet{Tokovinin_etal_2006} found that 96\% of a sample of
spectroscopic binaries with periods less than 3~days has a tertiary component.
Indeed, in some circumstances the distant companion enhances tidal interactions
in the inner binary, causing the binary orbital period to shrink to the
currently observed values. 

Three-body systems can be stable for long-time scales provided that the system
is hierarchical, that is, if the system is formed by an inner binary (star and
planet) in a nearly Keplerian orbit with a semi-major axis $ a $, and a outer
star also in a nearly Keplerian orbit about the center of
mass of the inner system  with semi-major $ a' \gg a $.
An additional requirement is that the eccentricity $ e' $ of the outer orbit is
not too large, in order to prevent close encounters with the inner system.
In this situation, perturbations on the inner planetary orbit are weak, but can
have important long-term effects (Chapter~10: {\it Non-Keplerian Dynamics}).

\begin{figure*}[t]
 \epsscale{2.}
\plotone{\figpath 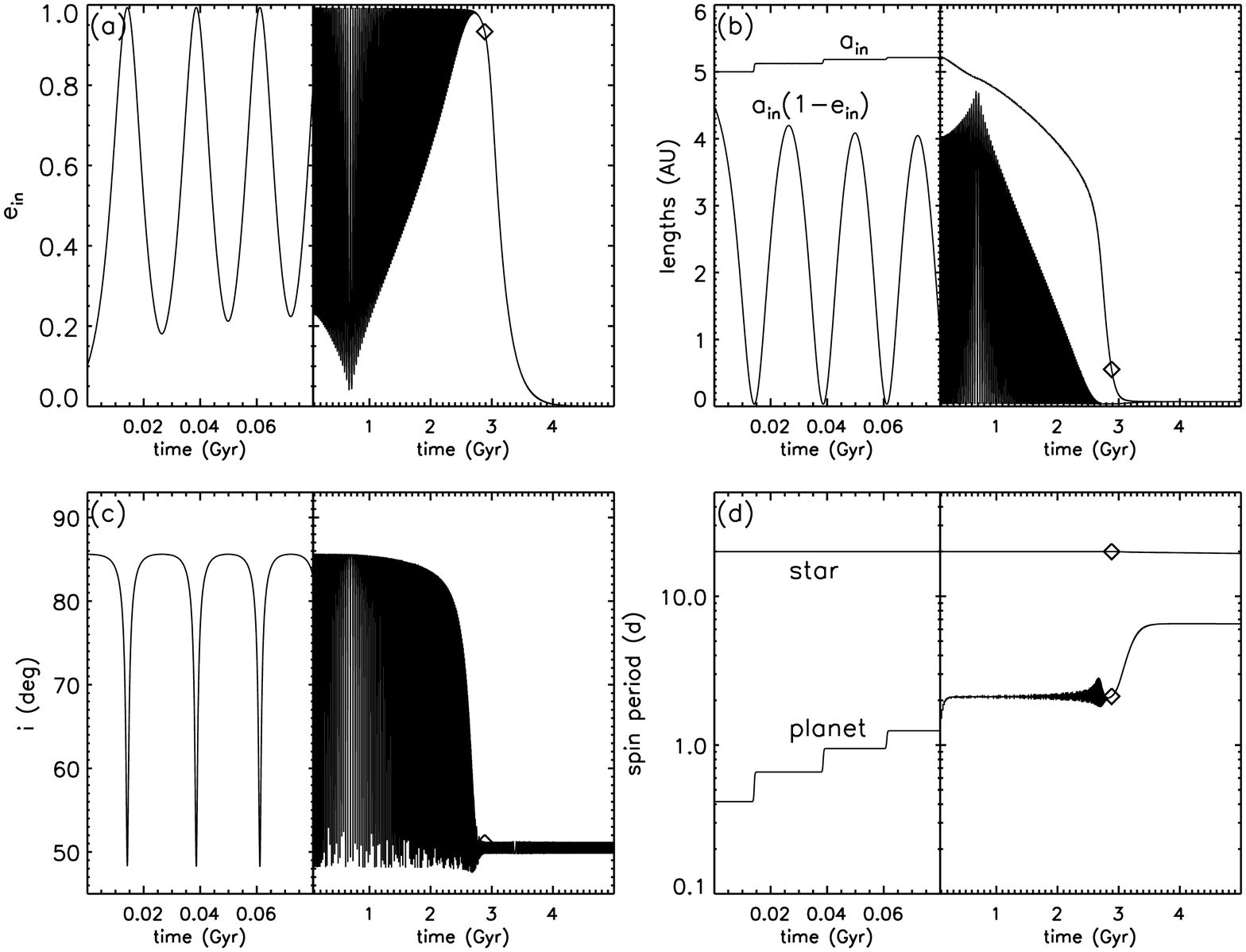}
  \caption{\small Possible evolution of the planet HD\,80606\,b initially in an
  orbit with $ a = 5$~AU, $ e = 0.1$, and $ I = 85.6^\circ $.
The stellar companion is supposed to be a Sun-like star at $ a' = 1000$~AU,
and $ e' = 0.5 $. The diamonds mark the current position of HD\,80606\,b along
this possible evolution \citep{Wu_Murray_2003,Fabrycky_Tremaine_2007}.
\llabel{kozai}}
 \end{figure*}

The most striking effect is known as the Lidov-Kozai mechanism
\citep{Kozai_1962,Lidov_1962}, which allows the inner orbit to periodically
exchange  eccentricity with inclination.
Even at large distances ($ a' > $1000~AU), the outer star can significantly
perturb the planetary orbit as long as the two orbital planes are initially
inclined to each other more than $I > 39.2^\circ$ (that is, for $\cos I <
(3/5)^{1/2}$).
When $I < 39.2^\circ$ there is little
variation in the planet's inclination and eccentricity.
Secular effects of the Lidov-Kozai type can then produce large cyclic
variations in the planet's eccentricity $e$ as a result of angular momentum
exchange with the companion orbit. 
Since the $z$-component of the planet's angular momentum must be conserved and
$a$ is not modified by the secular perturbations, the Kozai integral
\be
L_K = (1-e^2)^{1/2} \cos I \llabel{100211a}
\ee
is conserved during the oscillations  \citep{Lidov_Ziglin_1976}.
Maxima in $e$ occur with minima in $I$, and vice versa.
If the inner orbit is initially circular, the maximum eccentricity achieved in
a Kozai cycle is $ e_{max} = (1 - (5/3) \cos^2 I)^{1/2} $ and the oscillation
period of a cycle is approximately $ P'^2 / P $ \citep{Kiseleva_etal_1998}.
The maximum eccentricity of the inner orbit in the Kozai cycle will remain
fixed for different masses and distances of the outer star, but the period of
the Kozai cycle will grow with $a'^3$.
Kozai cycles persist as long as the perturbation from the outer star is the
dominant cause of periapse precession in the planetary orbit. 
However, small additional sources of periapse precession such as the quadrupole
moments, additional companions, general relativity or even tides can compensate
the Kozai precession and suppress the eccentricity/inclination oscillations
\citep[e.g.][]{Migaszewski_Gozdziewski_2009}.

Because the Lidov-Kozai mechanism is able to induce large eccentricity
excitations, a planet in an initial almost circular orbit (for instance a
Jupiter-like planet at 5~AU around a Sun-like star) 
can experiment close approaches to the host star at the periapse when the
eccentricity increases to very high values.
As a consequence, tidal effects increase several orders of magnitude and
according to expression (\ref{090522a}) the semi-major axis of the orbit will
decrease and the planet migrate inward.
At some point of the evolution, the periapse precession will be dominated by
other effects and the eccentricity oscillations suppressed.
From that moment on, the eccentricity is damped according to expression
(\ref{090522b}) and the final semi-major axis given by $ a_f = a (1-e^2) $.
\citet{Ford_Rasio_2006} have derived that tidal evolution of  high eccentric
orbits would  end  at a semi major axis $a_f$ equal to about twice the Roche
limit $R_L$. Indeed, at  the closest periapse distance, 
attained for $e\approx 1$, we will have  $a (1-e) = R_L$, 
and thus $a_f = (1+e) R_L \approx 2 R_L $.

In Figure~\ref{kozai} we plot an example of combined Kozai-tidal migration of
the planet HD\,80606\,b.
The planet is initially set in an orbit with $ a = 5$~AU, $ e = 0.1$, and $ I =
85.6^\circ $.
The stellar companion is supposed to be a Sun-like star at $ a' = 1000$~AU,
and $ e' = 0.5 $ \citep{Wu_Murray_2003,Fabrycky_Tremaine_2007}.
Prominent eccentricity oscillations are seen from the very beginning and the
energy in the planet's spin is transferred to the orbit increasing the semi-major
axis for the first 0.1~Gyr (Eq.\,\ref{090515a}).
As the equilibrium rotation is achieved (Eq.\,\ref{090520a}) the orbital evolution
is essentially controlled by equations (\ref{090522a}) and (\ref{090522b}),
whose contributions are enhanced when the eccentricity reaches high values.
The semi-major axis evolution is executed by apparent
``discontinuous'' transitions precisely because the tidal dissipation is only
efficient during periods of high eccentricity.
As dissipation shrinks the semi-major axis, periapse precession becomes
gradually dominated by relativity rather than by the third body, and the
periapse starts circulating as the eccentricity passes close to 0 at 0.7~Gyr. 
Tidal evolution stops when the orbit is completely circularized.
The present semi-major axis and eccentricity of planet HD\,80606\,b are $ a =
0.45$~AU and $ e = 0.92 $, respectively, meaning that the tidal evolution on
HD\,80606\,b is still under way (Fig.\,\ref{LC4}).
The final semi-major axis is estimated to about $ a_f = 0.07$~AU, which
corresponds to a regular ``Hot-Jupiter''.

\subsection{``Super-Earths''}

After a significant number of discoveries of gaseous giant exoplanets, a new
barrier has been passed with the detections of several exoplanets in the
Neptune and even Earth-mass ($M_{\oplus}$) regime: $2 - 12$ $M_{\oplus}$
\citep{Rivera_etal_2005,Lovis_etal_2006,Udry_etal_2007,Bonfils_etal_2007}, 
that are commonly designated by ``Super-Earths''.
If the commonly accepted core-accretion model can account
for the formation of ``Super-Earths'', resulting in a mainly icy/rocky
composition, the fraction of the residual He-H$_2$ atmospheric envelope accreted
during the planet migration is not tightly constrained for planets more massive
than the Earth \citep[e.g.][]{Alibert_etal_2006}. 
A minimum mass of below 10 $M_{\oplus}$ is usually considered to be the boundary between terrestrial
and giant planets, but \citet{Rafikov_2006} found that planets more massive than 6 $M_{\oplus}$ could have
retained more than 1 $M_{\oplus}$ of the He-H$_2$ gaseous envelope. For comparison, masses of Earth's
and Venus' atmosphere are respectively $\sim 10^{-6}$ and $10^{-4}$ times the planet's mass.
Despite significant uncertainties, the discoveries of ``Super-Earths'' provide
an opportunity to test some properties that could be similar to those of the
more familiar terrestrial planets of the Solar System.

Because some of the ``Super-Earths'' are potentially in the ``{\it habitable zone}''
\citep{Udry_etal_2007,Selsis_etal_2007}, the present spin state is an important
factor to constrain the climates. 
As for Venus, thermal atmospheric tides may have
a profound influence on the spin of ``Super-Earths''.
However, the small eccentricity approximation  done for Venus (Eq.\,\ref{eqSys})
 may no longer be adequate for  ``Super-Earths'',
which exhibit a wide range of eccentricities, orbital distances,
or central star types. 
Although our knowledge of ``Super-Earths''  is restricted to their orbital
parameters and minimum masses, we can attempt to place new constraints on the
surface rotation rate, assuming that ``Super-Earths'' have a dense atmosphere.

As for Venus, the combined
effect of tides is to set the final obliquity at $0^\circ$ or $180^\circ$
\citep{Correia_etal_2003}.
Adopting a viscous dissipation model for tidal effects (Eq.\,\ref{091126a}) and the 
``{\it heating at the ground}'' model \citep{Dobrovolskis_Ingersoll_1980} for
surface pressure variations (Eq.\,\ref{A11}), the average evolution of the
rotation rate is then obtained by adding the effects of both tidal
torques acting on the planet.
From expressions (\ref{021010c1}) and (\ref{A6a}) we get for $ \ve = 0^\circ $
and to the second order in the eccentricity:
\begin{eqnarray}
\frac{ \dot \omega}{\tau_{eq}^{-1}} = &  \! \! \! \! 
\omega - \left(1 + 6 e^2\right) n - \omega_s
\left[ \left(1 - \aa e^2\right) \sign (\omega - n)  \right.  \crm
& \! \! \! \! \! \! \! \! \! \! \! \! 
\left. \ab \, e^2 \sign (2 \omega - n) \ac \, e^2 \sign (2 \omega - 3 n)
\right] \ , \llabel{eq53}
\end{eqnarray}
where
\begin{equation}
\omega_s = \frac{F_s}{16 H_0 k_2} \frac{K_a \Delta t_a}{K_g \Delta
t_g} \propto \frac{L_\star}{m_\star} \, \frac{R}{m} \, a \ . \llabel{eq54}
\end{equation}

\begin{figure}[t]
 \epsscale{.87}
\plotone{\figpath 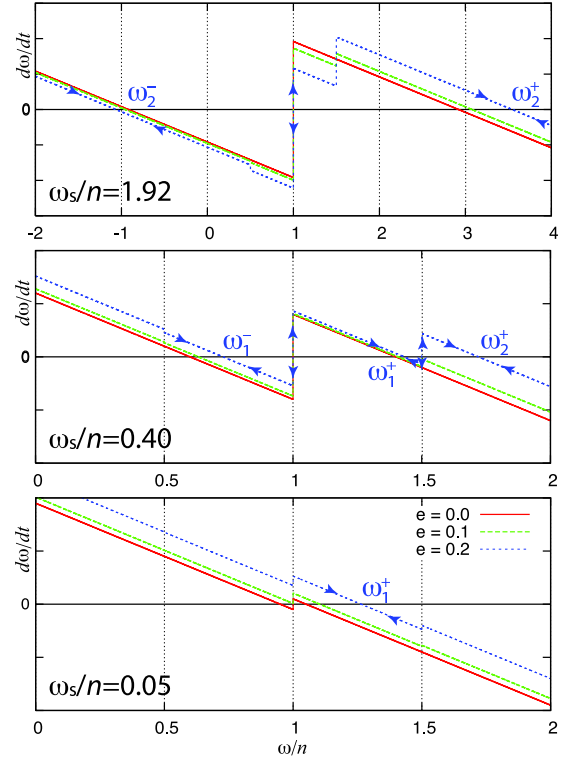}
  \caption{\small Evolution of $ \dot \omega $ (Eq.\,\ref{eq53}) with $ \omega /
  n $ for different atmospheric strengths ($ \omega_s / n = 1.92, 0.40, 0.05 $)
  and eccentricities ($ e = 0.0, 0.1, 0.2 $).
The top picture with $ e = 0 $ is the same as Figure~\ref{F6} for Venus.
The equilibrium rotation rates are given by $ \dot \omega = 0 $ and the arrows
indicate whether the equilibrium position is stable or unstable.
For $ \omega_s / n > 1 $, we have two equilibrium possibilities, $
\omega^\pm_2 $, one of which corresponds to a retrograde rotation (as
for Venus).
For $ \omega_s / n < 1 $, retrograde states are not possible, but we can still
observe final rotation rates $ \omega^- < n $.
For eccentric orbits, because of the harmonics in $ \sigma = 2 \omega - n $ and $
\sigma = 2 \omega - 3 n $, we may have at most four different final
possibilities (Eq.\,\ref{eq58a}).
When $ \omega_s /n $ becomes~extremely small, which is the case for the present
observed exoplanets with some eccentricity (Table\,\ref{Tab1}), a single
final equilibrium is possible for $ \omega^+_1 $ \citep{Correia_etal_2008L}. 
\llabel{Fig1}}
 \end{figure}

\begin{figure*}[t]
 \epsscale{2.}
\plotone{\figpath 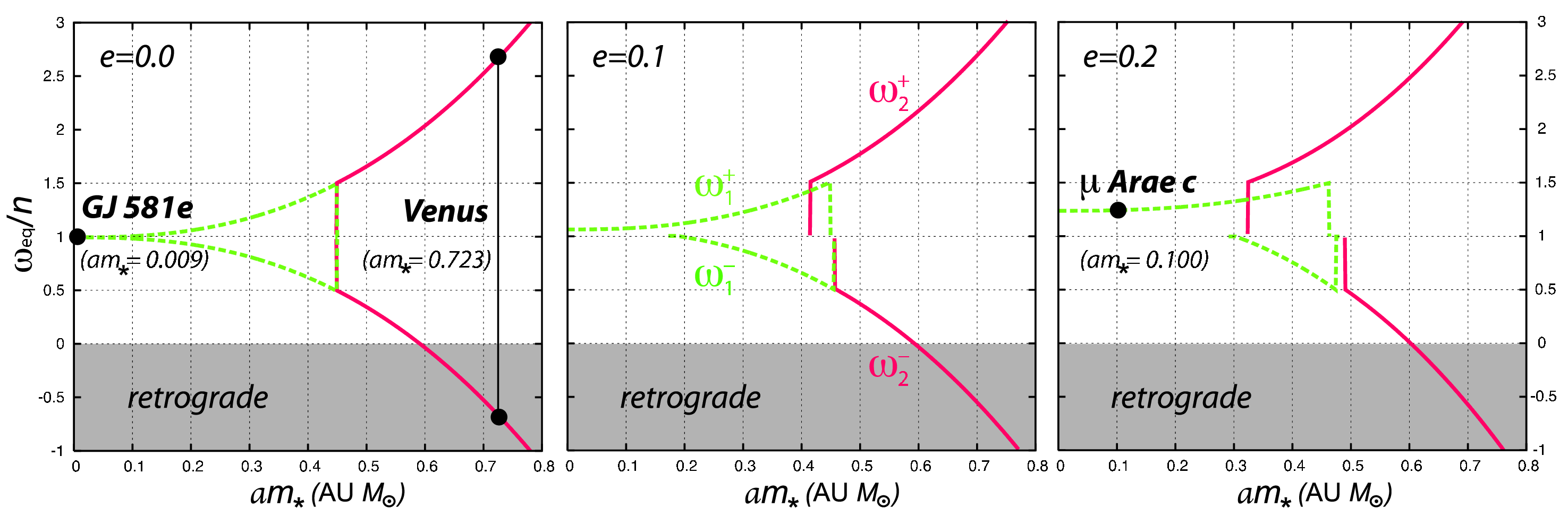}
  \caption{\small Equilibrium positions of the rotation rate for
  ``Super-Earths'' as a function of the product
$ a m_\star $ for three different values of the eccentricity ($e = 0.0, 0.1, 0.2 $).
Each curve corresponds to a different final state (dotted lines for $\omega_1^\pm
$ and solid lines for $\omega_2^\pm $).
For $ e \approx 0 $ (case of Venus), we always count two final states that are symmetrical
about $ n $. For small values of $ a m_\star $, the two equilibrium
possibilities are so close to $ n $ that the most likely scenario for the planet is to
be captured in the synchronous resonance (case of GJ\,581\,e).
As we increase the eccentricity, we can count at most three final equilibrium rotations,
depending on the value of $\omega_s/n$ (computed from Eq.\,\ref{eq60}).
When $ e \approx 0.2 $, only one equilibrium state exists for $ a m_\star < 0.3 $,
resulting from $ \omega_s / n < 6 e^2 ( 1 + e^2 / 2 ) $. This is 
the present situation of $\mu$\,Arae\,c and most of the ``Super-Earths''
listed in Table\,\ref{Tab1} \citep{Correia_etal_2008L}. \llabel{Fig2b}}
 \end{figure*}

\begin{deluxetable}{lcccccc|cccccc}
\tabletypesize{\small}
\tablecaption{Characteristics and equilibrium rotations of some ``Super-Earths''
with masses lower than $12 \, M_{\oplus}$ \llabel{Tab1}}
\tablewidth{0pt}
\tablehead{
 Name & $m_\star$ & Age & $\tau_{eq}$ & $ m \sin i $ & $a$ & $ e $ &
 $ \omega_s / n $ & $ 2 \pi /n $ & $ 2 \pi / \omega^-_2 $ &
 $ 2 \pi / \omega^-_1 $ & $ 2 \pi / \omega^+_1 $ & $ 2 \pi / \omega^+_2 $ \\

 & [$M_\odot$] & [Gyr] & [Gyr] & $[M_\oplus]$ & [AU]  & & &
 [day] & [day]  & [day]  & [day] & [day] \\
}
\startdata
Venus & 1.00 & 4.5 & 2.3 & 0.82 & 0.723 & 0.007 &
1.92 & 224.7 & $-$243.0   & &   & 76.8 \\ 

Earth$^{*}$ & 1.00 & 4.5 & 16 & 1.00 & 1.000 & 0.017 &
3.75 & 365.3 &  $-$132.9 &  &   & 77.1 \\ 

$^1$GJ\,581\,e  & 0.31 &7-11 & $10^{-7}$ & 1.94 & 0.03 & 0 &
 $10^{-5}$ & 3.4087 &   & 3.4088  & 3.4087 &   \\

$^2$HD\,40307\,b & 0.77 & ---  & $10^{-7}$ & 4.2 & 0.047 & 0 &
0.0003 & 4.2413 &  & 4.2427  & 4.240 &   \\

$^1$GJ\,581\,c & 0.31 & 7-11 & $10^{-5}$ &
5.36 & 0.07 & 0.17 &
$10^{-4}$ & 12.14 &   &   & 10.6335 &   \\

$^3$GJ\,876\,d & 0.32 & 9.9 &10$^{-8}$ & 6.3 & 0.021 & 0.14 &
$10^{-6} $ & 1.9649 &  &   & 1.7822 &   \\

$^2$HD\,40307\,c & 0.77 & ---  & $10^{-5}$ & 6.9 & 0.081 & 0 &
0.0009& 9.5956 &  & 9.6042  & 9.5871 &   \\

$^1$GJ\,581\,d & 0.31 & 7-11 & 0.02 & 7.09 & 0.22 & 0.38 &
0.0011 & 67.6918 &   &   & 47.8226 &   \\

$^4$HD\,181433\,b & 0.78 & --- & $10^{-5}$ & 7.5 & 0.08 & 0.396 &
0.0008 & 9.3579 &   &   & 6.535 &   \\

$^5$GJ\,176\,b & 0.5 & --- & $10^{-5}$ & 8.4 & 0.066 & 0 &
0.0002 & 8.7583 & &  8.7596  & 8.7568 &   \\

$^2$HD\,40307\,d & 0.77 & --- & $10^{-4}$ & 9.2 & 0.134 & 0 &
0.0025& 20.4175 &  & 20.4696  & 20.3656 &   \\

$^6$HD\,7924\,b & 0.83 & --- & $10^{-6}$ & 9.26 & 0.057 & 0.17 &
0.0004& 5.4493 &  &   & 4.7688 &   \\

$^7$HD\,69830\,b  & 0.86 & 4-10 & $10^{-5}$ & 10.2 & 0.079 & 0.10 &
0.0008 & 8.6625 &   &   & 8.1995 &   \\

$^8$$\mu$\,Arae\,c & 1.1 & 6.41 & $10^{-5}$ & 10.6 & 0.091 & 0.172 &
0.0021 & 9.5505 &   &   & 8.13313&   \\

$^9$55\,Cnc\,e  & 1.03 &5.5  & $10^{-7}$ & 10.8 & 0.038 & 0.07 &
0.0002 & 2.6659 &   &   & 2.592&   \\

$^{10}$GJ\,674\,b  & 0.35 & 0.1-1 & $10^{-7}$ & 11.09 & 0.039 & 0.2 &
$10^{-5} $ & 4.7549 &   &   & 4.0138 &   \\

$^7$HD\,69830\,c  & 0.86 & 4-10 & $10^{-3} $ & 11.8 & 0.186 & 0.13 &
0.0064 & 31.5943 &   &   & 28.8691 &   \\
\enddata
\smallskip

($^{*}$) Moon tidal effects were not included.
$\tau_{eq}$ was computed with $ k_2 = 1/3 $ and $ \Delta t_g = 640 $\,s
(Earth's values). 
References: [1] \citet{Mayor_etal_2009b}; [2] \citet{Mayor_etal_2009a};
[3] \citet{Correia_etal_2010}; [4] \citet{Bouchy_etal_2009};
[5] \citet{Forveille_etal_2009}; [6] \citet{Howard_etal_2009};
[7] \citet{Lovis_etal_2006}; [8] \citet{Pepe_etal_2007};
[9] \citet{Fischer_etal_2008}; [10] \citet{Bonfils_etal_2007}.
\end{deluxetable}

When $ e = 0 $, we saw in the case of Venus that final positions of the
rotation rate at zero obliquity are given by (Eq.\,\ref{eq7}):
\begin{equation}
| \omega - n | = \omega_s \ , \llabel{eq15}
\end{equation}
i.e., there are two final possibilities for the equilibrium rotation of the
planet, given by $ \omega^\pm = n \pm \omega_s $.
When $ e \ne 0 $, the above expression (\ref{eq15}) is no longer valid and additional equilibrium
positions for the rotation rate may occur.
For moderate values of the eccentricity, from expression (\ref{eq53}) we have
that the effect of the
eccentricity is eventually to split each previous equilibrium rotation rate into
two new equilibrium values. 
Thus, four final equilibrium positions for the
rotation rate are possible (eight if we consider the case $ \ve = 180^\circ $),
obtained with $ \dot \omega = 0 $ (Fig.\,\ref{Fig1}): 
\begin{equation}
\omega^\pm_{1,2} = n \pm \omega_s + e^2 \, \dw^\pm_{1,2} \ , \llabel{eq58a}
\end{equation}
with
\begin{equation}
\dw^-_{1} =  6 n + \frac{1}{2} \, \omega_s \ , \quad
\dw^+_{1} =  6 n - \frac{41}{2} \, \omega_s \ , \llabel{eq59a}
\end{equation}
and
\begin{equation}
\dw^-_{2} =  6 n + \frac{5}{2} \, \omega_s \ , \quad
\dw^+_{2} =  6 n - \frac{5}{2} \, \omega_s \ . \llabel{eq59b}
\end{equation}
Because the set of $\omega^\pm_{1,2}$ values must verify the additional condition
\begin{equation}
\omega^-_2 < n/2 < \omega^-_1 < n < \omega^+_1 < 3 n / 2 <  \omega^+_2 \ ,
\llabel{cond2}
\end{equation}
the four equilibrium rotation states cannot, in general, exist simultaneously,
depending on the values of $\omega_s$ and $ e $.
In particular, the final states $\omega^-_1$ and $\omega^+_1$
can never coexist with $ \omega^-_2 $. 
At most three different equilibrium states are therefore possible, obtained when
$\omega_s/n$ is close to $1/2$, or more precisely, when $1/2-19\,e^2/4 <
\omega_s / n < 1/2 + 17\,e^2/4$.
Conversely, we find that one single final state $\omega^+_1 = (1 + 6 e^2 ) \, n
+ (1 - 41 e^2 / 2) \, \omega_s $ exists when $\omega_s / n < 6 e^2 ( 1 + e^2 / 2)$.


The Earth and Venus are the only  planets that can be included in the
category of ``Super-Earths'' for which the atmosphere and spin
are known.
Only Venus is tidally evolved and therefore suitable for applying the above
expressions for tidal equilibrium.
We can nevertheless investigate the final equilibrium rotation states of
the already detected ``Super-Earths''.
For that purpose, we considered only exoplanets with masses smaller than
12\,$M_{\oplus}$ that we classified as rocky planets with a dense atmosphere,
although we stress that this mass boundary is quite arbitrarily.

Using the empirical mass-luminosity relation $ L_\star \propto M_\star^{4} $
\citep[e.g.][]{Cester_etal_1983} and the mass-radius
relationship for terrestrial planets $ R \propto m^{0.274} $ \citep{Sotin_etal_2007},
expression (\ref{eq54}) can be written as:

\begin{equation}
\omega_s / n = \kappa \, (a \, m_\star)^{2.5}  m^{-0.726} \ , \llabel{eq60}
\end{equation}
where $ \kappa $ is a proportionality coefficient that contains all the constant
parameters, but also the parameters that we are unable to
constrain such as $ H_0 $, $ k_2 $, $ \Delta t_g $ or $ \Delta t_a $.
In this context, as a first order approximation 
we consider that for all ``Super-Earths'' the parameter $ \kappa $ 
has the same value as for Venus.
Assuming that the rotation of Venus is presently stabilized in the $ \omega^- $
final state, that is, $ 2 \pi / \omega^- = - 243 $\,days
\citep{Carpenter_1970}, we compute $ 2 \pi / \omega_s = 116.7 $\,days.
Replacing the present rotation in expression (\ref{eq60}), we find for Venus that
$ \kappa = 3.723 \; M_\oplus^{0.726} M_\odot^{-2.5} $AU$^{-2.5}$.
We can then estimate the ratio $ \omega_s/n $ for all considered ``Super-Earths'' 
in order to derive their respective equilibrium rotation rates (Table\,\ref{Tab1}).

The number and values of the allowed equilibrium rotation states are plotted as a function
of $a M_\star $ for different eccentricities in Figure\,\ref{Fig2b}.
All eccentric planets have a ratio $\omega_s/n$ that is lower than $6 \times
10^{-3}$ (Table\,\ref{Tab1}),
which verifies the condition  $ \omega_s / n < 6 e^2 (1 + e^2 / 2) $. As a consequence,
only one single final state exists $\omega^+_1 / n \approx (1+6\,e^2) $, corresponding to the equilibrium
rotation resulting from gravitational tides (Eq.\,\ref{090520a}). 
The main reason
is that the effect of atmospheric tides is clearly disfavored relative to the
effect of gravitational tides on ``Super-Earths'' discovered orbiting
M-dwarf stars:
the short orbital periods strengthens the effect of gravitational tides, which
are proportional to $1/a^6$, while the effect of thermal tides varies as $1/a^5$. 
Moreover, the small mass of the
central star also strongly affects the luminosity received by the planet and hence
the size of the atmospheric bulge driven by thermal contrasts.

For the planets with nearly zero eccentricity (GJ\,581\,e,
HD\,40307\,b,\,c,\,d, and GJ\,176\,b),
two equilibrium rotation states $\omega^\pm_1 $ are possible. 
However, the two final states $\omega^\pm_1 $ are so 
close to the mean motion $ n $, that the quadrupole moment of inertia $(B-A)/C$
will probably capture the rotation of the planet in the synchronous resonance.
We then conclude that ``Super-Earths'' orbiting close to their
host stars (in particular M-dwarfs), will be dominated by gravitational tides
and present a final equilibrium rotation rate given by $ \omega_e / n \approx f_2
(e) / f_1(e) $ (Fig.\,\ref{FigC}), or present spin-orbit resonances like
Mercury. 

\section{FUTURE PROSPECTS}


The classical theory of tides initiated by \citet{Darwin_1880,Darwin_1908} is
sufficient to understand the main effects of tidal friction upon planetary
evolution. 
However, the exact mechanism on how tidal energy is dissipated within the
internal layers of the planet remains a challenge for planetary scientists.
\citet{Kaula_1964} derived a generalization of Darwin's work, with
consideration of higher order tides and without the adoption of any dissipation
model.
The tidal potential is described using infinite series in eccentricity and
inclination, which is not practical and can only be correctly handled by
computers.
Ever since many efforts have been done in order to either simplify the tidal
equations, or to correctly model the tidal dissipation 
\citep[for a review see][]{Ferraz-Mello_etal_2008,Efroimsky_Williams_2009}.

Many Solar System phenomena have been successfully explained using the existent
tidal models, so we expect that they are suitable to describe the tidal
evolution of exoplanets.
Nevertheless, many exoplanets are totally different from the Solar
System cases, and we cannot exclude to observe some unexpected behaviors.
For instance, it is likely that dissipation within ``Hot-Jupiters'' is closer to
dissipation within stars \citep[e.g.][]{Zahn_1975}, while dissipation within
``Super-Earths'' is closer to dissipation observed for rocky planets
\citep[e.g.][]{Henning_etal_2009}.
It is then necessary to continue improving tidal models in order to get a more
realistic description for each planetary system.
In particular, a correct description of the tidal dissipation and on how it
evolves with the tidal frequency is critical for the evolution time-scale.

The orbital architecture of exoplanetary systems is relatively well determined
from the present observational techniques. 
However, the spins of the exoplanets are not easy to measure, as the light curve
coming from the planet is always dimmed by the star light.
The continuous improvements that have been made in photometry and
spectrography let us believe that the determination of exoplanets' spins can be
a true possibility in the near future. 
In particular, infra-red spectrographs are being developed, which will allow to
acquire spectra of the planets if we manage to subtract the stellar
contribution \citep[e.g.][]{Barnes_etal_2010}.

Some additional methods for detecting the rotation and/or the obliquity of
exoplanets have also been tested and suggested so far.
For instance, indirect sensing of the planetary gravitational quadrupole
and shape, which is linked to both spin rate and obliquity
\citep[e.g.][]{Seager_Hui_2002,Ragozzine_Wolf_2009}, or 
transient heating of one face of the planet, which then spins
into and out of view, as it has been attempted for the system HD\,80606
\citep{Laughlin_etal_2009}.
The effect of planetary rotation on the
transit spectrum of a giant exoplanet is another possibility. During ingress and egress,
absorption features arising from the planet's atmosphere are Doppler shifted by
of order the planet's rotational velocity ($\sim$1-2\,km\,s$^{-1}$) relative to where they
would be if the planet were not rotating \citep[e.g.][]{Spiegel_etal_2007}. 
Finally, for planets whose light is spatially separated from the star,
variations may be discernible in the light curve obtained by low-precision
photometry due to meteorological variability, composition of the surface, or
spots \citep[e.g.][]{Ford_etal_2001N}.

Although the spin states of exoplanets cannot be measured, for exoplanets that
are tidally evolved we can still try to make predictions for the rotation rates.
When the eccentricity is large, the rotation of many of the observed exoplanets
can still be tidally evolved even if the planets are not very close to their 
central stars (Fig.\,\ref{LC3}). 
For tidally evolved ``Hot-Jupiters'', we can conjecture that the rotation
periods are the limit values $P_{orb} \times f_1(e) / f_2(e) $ (Fig.\,\ref{FigC}). 
It becomes a new challenge for the observers to be able to confirm these
predictions. 

Thermal atmospheric tides may very well destabilize the tidal equilibrium from
gravitational tides and create additional possible stable limit values, with the
possibility of retrograde rotations, as for planet Venus (Fig.\ref{F6}).
Thermal tides should be particularly important for ``Super-Earths'', which are
expected to have a distinct rocky body surrounded by a dense atmosphere.
In a paradoxical way, the final rotation rate of ``Super-Earths'' are the most
difficult to predict, as the equilibrium configurations depend on the
composition of the atmospheres.  
Thermal tides are nevertheless more relevant for exoplanets that orbit Sun-like
stars at not very close distances, like Venus (Fig.\ref{Fig2b}).

We also assumed that the final obliquity of exoplanets is
either $0^\circ$ or $180^\circ$, as the two values represent the final
outcome of tidal evolution.
However, each planetary system has its own architecture, and planetary
perturbations on the spin can lead to resonant capture in a high oblique Cassini
states or even to chaotic motion.
Thus, the final spin evolution of a planet cannot be dissociated from its
environment, and a more realistic description of exoplanets rotation
can only be achieved with the full knowledge of the system orbital dynamics.

\bigskip
\textbf{ Acknowledgments.} We thank to an anonymous referee for valuable
suggestions, who helped to improve this work.
We acknowledge support from the Funda\c{c}\~ao para a Ci\^encia e a Tecnologia
(Portugal) and PNP-CNRS (France).

\bigskip

\parskip=0pt

\bibliographystyle{exoplanets}
\bibliography{correia}

\end{document}